\documentclass[preprint]{vgtc}               

\vgtcinsertpkg

\ifpdf
  \pdfoutput=1\relax                   
  \usepackage{graphicx}                
  \DeclareGraphicsExtensions{.pdf,.png,.jpg,.jpeg} 
\else
  \usepackage{graphicx}                
  \DeclareGraphicsExtensions{.eps}     
\fi%

\graphicspath{{figures/}{pictures/}{images/}{./}} 

\usepackage{url}
\usepackage[utf8]{inputenc}
\usepackage[T1]{fontenc}
\usepackage{microtype}                 
\PassOptionsToPackage{warn}{textcomp}  
\usepackage{textcomp}                  
\usepackage{mathptmx}                  
\usepackage{times}                     
\usepackage{cite}                      
\usepackage{booktabs}                  
\usepackage{amsmath}
\usepackage{multirow}
\usepackage{cleveref}
\usepackage{t1enc}
\usepackage{tabularx}
\usepackage{makecell}
\usepackage{xspace}
\usepackage[dvipsnames]{xcolor}
\usepackage{soul}
\usepackage{color}
\usepackage{relsize}
\usepackage[plain]{algorithm}
\usepackage[noend]{algpseudocode}
\usepackage[labelfont=sf]{subcaption}
\usepackage[frozencache=true,cachedir=./]{minted}
\definecolor{mintedbgcolor}{rgb}{0.95,0.95,0.95}
\usepackage{xcolor}
\usepackage{tcolorbox}
\usepackage{minted}
\setminted{fontsize=\fontsize{7}{7},mathescape}
\BeforeBeginEnvironment{minted}{
    \begin{tcolorbox}[colback=mintedbgcolor,left=1mm]
}
\AfterEndEnvironment{minted}{\end{tcolorbox}}

\onlineid{1012}

\vgtccategory{Research}

\title{Interactive Visualization of Terascale Data in the Browser:  \\ 
Fact or Fiction?}

\author{Will Usher\thanks{will@sci.utah.edu} \;\;\;\;\;\; Valerio Pascucci \\
\parbox{3.4in}{\scriptsize \centering SCI Institute, University of Utah}}

\teaser{
  \vspace{-1em}
  \includegraphics[width=\textwidth]{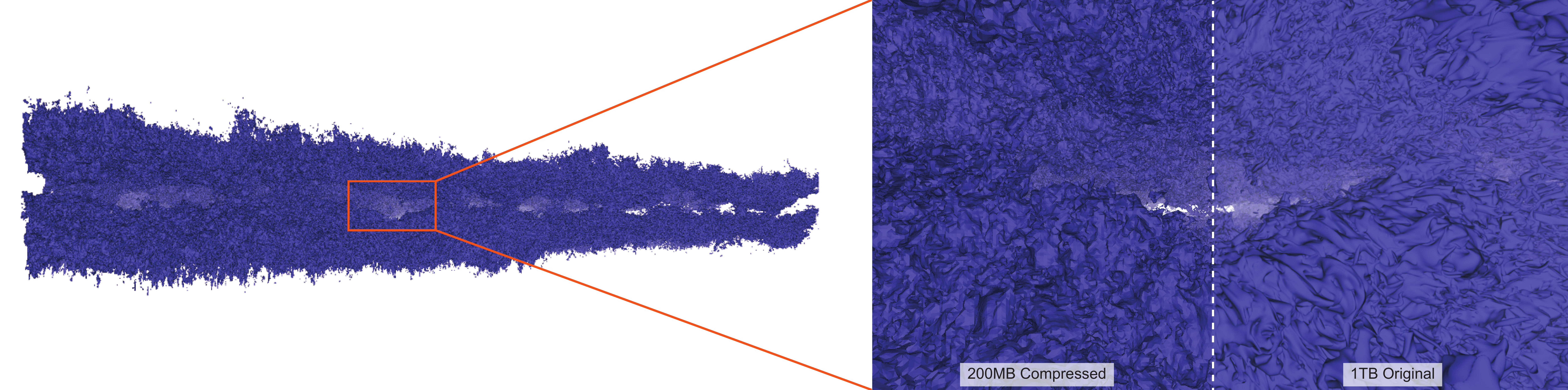}
  \vspace{-2em}
    \caption{\label{fig:teaser}%
    Interactive visualization of an isosurface of a \textasciitilde1TB dataset entirely in the web browser.
    The full data is a float64 10240$\times$7680$\times$1536 grid computed by a DNS simulation~\cite{dns_sc,dns_jfm}.
    The isosurface is interactively computed and visualized entirely in the browser using our GPU isosurface
    computation algorithm for block-compressed data, 
    after applying advanced precision and resolution trade-offs~\cite{hoang_efficient_2021,hoang-precision-tvcg-2019,lindstrom_fixed_rate_2014}.
    The surface consists of 137.5M triangles and is computed in 526ms on an RTX 2070 using WebGPU in Chrome
    and rendered at 30FPS at 1280$\times$720.
    The original surface, shown in the right split image, consists of 4.3B triangles and was computed
    with VTK's Flying Edges filter~\cite{schroeder_flying_2015} on a quad-socket Xeon server in 78s using 1.3TB of memory.}

}

\abstract{%
    Information visualization applications have become ubiquitous, in no
    small part thanks to the ease of wide distribution and deployment to
    users enabled by the web browser. Scientific visualization applications,
    relying on native code libraries and parallel processing,
    have been less suited to such widespread distribution, as browsers
    do not provide the required libraries or compute capabilities.
    In this paper, we revisit this gap in visualization technologies and explore
    how new web technologies, WebAssembly and WebGPU,
    can be used to deploy powerful visualization solutions 
    for large-scale scientific data in the browser.
    In particular, we evaluate the programming effort required to bring
    scientific visualization
    applications to the browser through these technologies and assess
    their competitiveness against classic native solutions.
    As a main example, we present a new GPU-driven isosurface
    extraction method for block-compressed data sets,
    that is suitable for interactive isosurface computation
    on large volumes in resource-constrained environments, such as the browser.
    We conclude that web browsers are on the verge of becoming a
    competitive platform for even the most demanding scientific visualization
    tasks, such as interactive visualization of isosurfaces from a 1TB DNS simulation.
    We call on researchers and developers to consider investing
    in a community software stack to ease use of these upcoming browser
    features to bring accessible scientific visualization to the browser.
}

\CCScatlist{
    Web Applications; Parallel Isosurface Extraction
}

\begin{document}


\firstsection{Introduction}

\maketitle

Information visualization applications using D3~\cite{bostock_d_2011},
Vega~\cite{satyanarayan_declarative_2014}, Vega-Lite~\cite{satyanarayan_vega_lite_2017},
and Tableau (formerly Polaris)~\cite{stolte_polaris_2002}
have become ubiqituous on the web, reaching millions of users to
make data more accessible and understandable.
The web browser platform is key to enabling this ubiquity,
by providing a standardized, cross-platform environment through which
applications can be deployed to users.
Similarly, the browser eases the process by which users access
an application and enhances security, removing the need to install
and trust additional packages on their system.
Although the browser has been widely used to
deploy information visualization applications, this has not been
the case for scientific visualization applications.


Scientific visualization applications often rely on native
libraries (e.g., VTK~\cite{vtk}), fast floating point operations,
large amounts of memory, and parallel processing or high-performance
rendering on the GPU or CPU. Porting the required
native libraries and rewriting the application to JavaScript is a
significant undertaking, and does not address the computational
demands of the application.
As a result, the majority of web-based scientific visualization
applications rely on a backend server to perform data processing
and visualization (e.g.,~\cite{raji_scientific_2018,raji_scalable_2017,perez_ipython_2007,
jourdain_paraviewweb_2011,dykes_interactive_2018}),
or to stream reduced representations
of the data to reduce the demands placed on the browser (e.g.,~\cite{schutz_potree_2016,jacinto_web_2012,
mobeen_high_performance_2012,sherif_brainbrowser_2015}).
Using a server gives the application
access to significant computational resources, at the cost of
latency for users and financial cost to the developer to run the
servers. When considering a large-scale deployment, the cost of
provisioning sufficient servers to meet demand is a real concern.

WebAssembly~\cite{webasm_web} and WebGPU~\cite{webgpu_web} are new web technologies that
can help address the issues faced when bringing scientific visualization
applications to the browser. WebAssembly is a bytecode format to which native
code can be compiled  (e.g., through Emscripten~\cite{zakai_emscripten_2011}) and that can be executed
at near-native speeds in the browser.
WebGPU is a new low-level graphics API for the browser, similar in
spirit to DirectX~12, Vulkan, and Metal, although it provides an easier-to-use API.
WebGPU exposes advanced
rendering and compute functionality to significantly expand the GPU
capabilities available to browser applications compared to WebGL.
WebGPU supports compute shaders and storage buffers,
enabling arbitrary parallel computation on large data,
and easing the implementation of such computation compared
to prior efforts, which leveraged the WebGL rendering pipeline~\cite{li_p4_2018,li_p5_2019}.


In this paper, we perform an initial evaluation of how these new
technologies can be leveraged to enable scientific visualization applications
to run directly in the browser. We evaluate the performance of widely
used native libraries for importing images and LIDAR data when compiled
to WebAssembly, and the performance of Marching Cubes~\cite{lorenson_marching_1987}
as proxy for common floating point and memory intensive scientific visualization algorithms.
Finally, we propose a fully GPU-driven parallel isosurface extraction algorithm
for interactive isosurfacing of block-compressed volume data on memory-constrained
client devices to enable large-scale data visualization in the browser.
Our isosurface extraction algorithm makes use of
GPU decompression and a GPU-managed LRU cache to achieve interactive
isosurface extraction with a small memory footprint.
Our specific contributions are:
\begin{itemize}
    \item An experimental assessment of WebAssembly and WebGPU as tools for building scalable scientific
        visualization applications, e.g., volume rendering, geospatial visualization, and isosurface
        computation, that execute entirely in the browser client without offloading computation
        to a server.
    \item A parallel isosurface extraction algorithm for resource-constrained platforms, such as the browser.
        The approach includes:
\begin{itemize}
\item A GPU-driven parallel isosurface extraction algorithm capable of interactive isosurface computation for interactive data interrogation;
\item A GPU-driven memory management and caching strategy that allows working with compressed data on the GPU without CPU interaction;
\item Evaluation of our approach on up to 1B voxels and 1TB of volume data by applying state-of-the-art precision and resolution trade-offs.
\end{itemize}
\end{itemize}

\section{Related Work}

Even with the browser's prior limitations on
graphics and compute capability, a number of works have explored
methods for deploying scientific visualization in the browser,
either through a client-server architecture or by leveraging
the capabilities provided by WebGL. We review these works
in~\Cref{sec:related_webscivis}. In~\Cref{sec:related_mc}
we review relevant work on parallel isosurface extraction algorithms.

\subsection{Scientific Visualization in the Browser}
\label{sec:related_webscivis}

A number of scientific visualization applications have used WebGL, a subset of OpenGL
available in the browser, to perform rendering on the client~\cite{schutz_potree_2016,jacinto_web_2012,
mobeen_high_performance_2012,sherif_brainbrowser_2015}.
Mobeen and Feng~\cite{mobeen_high_performance_2012} demonstrated a
volume renderer in WebGL capable of high-quality interactive volume
rendering of small data sets on mobile devices. Sherif et al.~\cite{sherif_brainbrowser_2015}
presented BrainBrowser to allow interactive visualization of large-scale neuroimaging
data sets fetched from a remote server. Recently, Li and Ma~\cite{li_p4_2018,li_p5_2019}
proposed an approach for parallel processing by leveraging the WebGL rendering pipeline
to provide a subset of common data-parallel primitives.
However, WebGL lacks support for compute shaders and storage buffers, making the implementation
of general high-performance parallel primitives challenging.

To visualize large data sets, prior work has either streamed subsets of
the data to clients for rendering~\cite{schutz_potree_2016,sherif_brainbrowser_2015}, or
rendered it remotely on a server and streamed the resulting images back~\cite{raji_scientific_2018,
raji_scalable_2017,perez_ipython_2007,jourdain_paraviewweb_2011,dykes_interactive_2018}.
Remote rendering approaches allow the application access to arbitrary compute capabilities
to process massive data sets, although these approaches can face challenges with latency and cost.
When considering a large-scale deployment, e.g., data visualization
on the home page of the New York Times,
the cost of running sufficient servers to satisfy the large number of clients
is a real concern. Fully client-side applications, our focus in this paper,
move the computation and data to the clients.
Users can then interact with the data locally, without added
network latency.

In a recent survey of web-based visualization approaches, Mwalongo et al.~\cite{mwalongo_stateart_2016}
specifically mention latency and the lack of CUDA capabilities as key remaining challenges
for visualization in the browser.
By using WebGPU, applications that rely on GPU computing can now be run
locally on the client, eliminating the cost of GPU virtual machines and
latency. Access to powerful GPU capabilities can also ease the implementation
of more advanced latency hiding techniques for hybrid client-server applications.

\subsection{Parallel Isosurface Extraction}
\label{sec:related_mc}
Isosurface extraction is a classic and widely used scientific visualization
algorithm, and has received wide attention since the original Marching Cubes paper~\cite{lorenson_marching_1987}.
Livnat et al.~\cite{livnat_near_1996} and Cignoni et al.~\cite{cignoni_optimal_1996}
proposed to accelerate computation by skipping
regions that are known to not contain the isosurface based on their range, using
a range k-d tree or interval tree to accelerate the search.
Approaches leveraging similar ideas for skipping regions, based on macrocell grids~\cite{parker_interactive_1998}
and k-d trees~\cite{wald_faster_2005,gross_fast_2007}, have been proposed
for implicit isosurface ray tracing, allowing the application to render
the surface directly without storing additional vertex data.

A number of GPU parallel isosurface extraction algorithms have been
proposed~\cite{dyken_highspeed_2008,ciznicky_efficient_2012,jeong_configurable_2012,
martin_loadbalanced_2010,schmitz_efficient_2010,
schroeder_flying_2015,liu_parallel_2016,cuda_mc_example}. GPU-parallel
algorithms typically process each voxel in parallel in a thread,
using prefix sums and stream compaction operations to output the vertices
produced into a single buffer. Early work achieved the
compaction step using HistoPyramids~\cite{dyken_highspeed_2008}
and performed processing in the geometry shader; however,
recent works~\cite{cuda_mc_example,liu_parallel_2016} in CUDA are able to leverage
compute kernels for processing and Thrust~\cite{bell_thrust_2017} to provide fast
prefix sum and stream compaction kernels. In contrast to our approach,
existing algorithms assume the entire data set is able to fit
in the memory of one GPU~\cite{dyken_highspeed_2008,ciznicky_efficient_2012,jeong_configurable_2012,
martin_loadbalanced_2010,schmitz_efficient_2010,schroeder_flying_2015,liu_parallel_2016,cuda_mc_example},
or be distributed among multiple GPUs in a cluster~\cite{martin_loadbalanced_2010}

The block-based work decomposition approach proposed by Liu et al.~\cite{liu_parallel_2016}
is most similar to our proposed algorithm. Liu et al. compute the value
range of each block
and use this information to filter out blocks that do not contain the isovalue.
They then compute the number of vertices to be output by each block in parallel and perform
a prefix sum to compute the output offsets for each block. A second
pass computes and outputs the vertices for each voxel. Per voxel output
offsets are computed using a prefix sum within the thread group processing
a block, eliminating the need to store per voxel offsets or perform a global prefix sum.
Within each block, vertices are computed using Flying Edges~\cite{schroeder_flying_2015}.
However, Liu et al. assume the entire data set is stored in a single 3D texture
and only use the block decomposition to accelerate computation. In contrast, our approach
does not store the full volume data. Instead, we decompress and cache only the blocks
required to compute the surface using our GPU-driven LRU cache.

%
%
%
%
%

\section{Revisiting the Browser as a Platform for Accessible Scientific Visualization}  

In this work, we are concerned with the capabilities available in
the user's browser and the applications that can be developed
by leveraging those capabilities.
There are two recent technologies in the browser that can be used
to develop powerful visualization applications:
WebAssemby (\Cref{sec:tech_wasm}) and WebGPU (\Cref{sec:tech_webgpu}).


\subsection{WebAssembly}
\label{sec:tech_wasm}

When considering bringing an existing scientific visualization application
to the web, or deploying an entirely new one, a major concern is the lack
of libraries widely used in native applications, e.g., for loading different data formats or
performing computation. Previously, developers may have chosen to port
the required libraries to JavaScript (e.g., VTK.js~\cite{vtkjs}),
or recompile them to JavaScript using Emscripten~\cite{zakai_emscripten_2011}.
Porting a substantial
set of dependencies to JavaScript is clearly a major undertaking,
and although Emscripten can compile C and C++ code to JavaScript,
the performance of JavaScript is typically insufficient for computationally
intensive tasks. Today, the above issues
can be addressed by using Emscripten, or toolchains for other languages,
to compile the required libraries to WebAssembly.

WebAssembly (Wasm)~\cite{webasm_web} is a standardized bytecode format for a
virtual machine that can be run in the browser or natively.
In contrast to Java Applets or Flash, WebAssembly is
integrated natively into the browser, supporting tight integration
with JavaScript and security through sandboxing.
and it is targetable by C and C++ through Emscripten.
WebAssembly is shipping today in the major browsers,
Chrome, Firefox, Safari, and Edge, ensuring wide availability on
user systems. In a study of various compute
benchmarks, Jangda et al.~\cite{jangda_not_2019} found WebAssembly
to be 1.5--3.4$\times$ slower than native code, depending
on the task. These slowdowns can be partly attributed to enforcing WebAssembly's
security guarantees, which require some additional checks to be performed.

\begin{listing}[t]
    \begin{minted}{c++}
struct LASFile {
  LASreader *reader;
  std::vector<float> positions;
  LASFile(const char *fname);  // Reads file using LAStools
};
extern "C" LASFile* openLAS(const char *fname) {
  return new LASFile(fname);
}
extern "C" float* getPositions(LASFile *file){
  return file->positions.data();
}
extern "C" uint64_t getNumPoints(LASFile *file) {
  return file->positions.size() / 3;
}
    \end{minted}
    \begin{minted}{javascript}
// Setup call information for the exported C functions
var openLAS = Module.cwrap("openLAS", "number", ["string"]);
var getPositions =
    Module.cwrap("getPositions", "number", ["number"]);
var getNumPoints =
    Module.cwrap("getNumPoints", "number", ["number"]);
// Write file data into Emscripten's virtual filesystem
FS.writeFile("data.laz", new Uint8Array(fileData));
var lasFile = openLAS("data.laz");
// Create a view of the Wasm module's point data
var positions = new Float32Array(HEAPF32.buffer,
    getPositions(lasFile), getNumPoints(lasFile) * 3);
    \end{minted}
    \vspace{-1em}
    \caption{\label{lst:liblas_mem_sharing}%
    A subset of our C API (top) for the LASlib WebAssembly module
    and its use from JavaScript (bottom).
    The C API manages loading the data using LASlib and stores
    it in memory that can be shared with JavaScript.
    JS creates a view of the module's memory
    starting at the ``pointer'' returned by the API
    to access the data.
    \vspace{-2em}}
\end{listing}

\subsubsection{Compiling Native Libraries to WebAssembly}
The Emscripten compiler can be used as a drop in replacement
C or C++ compiler and can compile most portable
code to WebAssembly\footnote{\href{https://emscripten.org/docs/porting/guidelines/index.html}{https://emscripten.org/docs/porting/guidelines/index.html}}.
The main challenge when compiling a library to WebAssembly is
the number of dependencies required by the library, as each one must
also be compiled to WebAssembly.
Large libraries such as VTK~\cite{vtk}, whose dependencies
may themselves have a number of dependencies, can produce a large
graph of libraries that must all be compiled to WebAssembly to link against.

To provide a convenient API for calling into C++ libraries from JavaScript and working
with C++ objects, library authors can use the binding generation tools provided by
Emscripten: WebIDL Binder (similar to SWIG) or Embind (similar to Boost.Python).
For smaller libraries or minimal bindings, it is also possible to export a C
API and call it directly from JavaScript.

We discuss two applications that use native
libraries compiled to WebAssembly. The Neuron Visualization example (\Cref{sec:example_neuron}),
which uses the C libtiff library to load TIFF image stacks, and the LIDAR visualization
example, which uses the C++ LASlib~\cite{lastools} library to load las and laz LIDAR data files.
The libtiff library depends on zlib and libjpeg, and thus we had to first compile
these libraries to Wasm and link them with libtiff to produce the library.
For LASlib, we wrote a minimal C API (\Cref{lst:liblas_mem_sharing}) to
discuss the lower level details of calling between JavaScript and C or C++ Wasm
modules.
Our libtiff\footnote{\href{https://github.com/Twinklebear/tiff.js}{https://github.com/Twinklebear/tiff.js}}
and LASlib\footnote{\href{https://github.com/Twinklebear/LAStools.js}{https://github.com/Twinklebear/LAStools.js}}
WebAssembly modules are available on Github.

%

\subsubsection{Sharing Memory Between JavaScript and WebAssembly Modules}
\label{sec:wasm_share_mem}

The heap space of a WebAssembly module is stored within a JavaScript
ArrayBuffer object that can be grown to meet dynamic allocation needs
of the module (i.e., \texttt{malloc} and \texttt{new}).
Pointers in the module are simply offsets into this ArrayBuffer.
Thus, the module can directly share memory with the JavaScript host
code by returning a pointer that can be used to create a
view of the module's heap starting at the returned index.
We use this approach to share the point cloud data loaded by our LASlib
module with JavaScript without copying (\Cref{lst:liblas_mem_sharing}).

However, it is not possible for WebAssembly modules to directly see JavaScript
memory. To pass an array of data from JavaScript to a module, space must be
allocated from the module's heap to copy the array into.
The index to which the array was written in the heap is then passed to the module
as a pointer.
To minimize the number
of such copies that must be made to pass arrays to the module, it is best
to have the module store large arrays in its heap
and have JavaScript create an array view of the heap.

\subsection{WebGPU}
\label{sec:tech_webgpu}


WebGPU~\cite{webgpu_web} is a modern graphics API for the web, in development by the
major browser vendors and available for testing in the preview builds
of Chrome, Firefox, and Safari. WebGPU fills a role similar to Vulkan, DirectX~12, and Metal,
providing a low-level API with more explicit control to the user and
fixed inputs to the driver, allowing for improved performance compared to WebGL.
Moreover, WebGPU exposes additional GPU functionality that is key to implementing
general parallel compute algorithms, providing access
to compute shaders, storage buffers, and storage textures.

The concepts used in WebGPU should be familiar
to users of Vulkan, DirectX~12, and Metal, though WebGPU is not
as ``low level'' as the native APIs, striking a good balance
between usability and performance. In WebGPU, rendering or compute passes are
recorded using a command encoder to buffer up and submit work to the GPU.
When recording a pass, a rendering or compute pipeline can be bound.
The pipeline specifies the shaders to run, the layout of buffer inputs to the shaders,
and, for rendering pipelines, the layout of vertex attributes and output render targets.
Groups of buffers and textures matching the specified layouts can be bound
to pass data to the shaders. The fixed layout of the pipeline allows the GPU
to optimize its execution while the data being processed can still be changed
as needed. This fixed layout is in contrast to WebGL and OpenGL, where the entire rendering
or compute state cannot be provided to the driver in advance, limiting the optimizations
that can be performed.


\subsection{Example Applications and Performance vs.\ Native Execution}

In this section, we evaluate a set of example applications for common visualization
tasks that make use of WebAssembly and WebGPU. The applications cover a range
of common operations performed in visualization: loading TIFF
image stacks of microscopy volumes (\Cref{sec:example_neuron}),
interactive visualization of LIDAR data (\Cref{sec:example_lidar}),
naive serial and data-parallel isosurface extraction (\Cref{sec:example_mc}),
and volume decompression (\Cref{sec:example_zfp}).
For each example we provide a performance comparison against native code
on the same task, to assess the capabilities of WebAssembly and WebGPU
for scientific visualization applications.

\subsubsection{Neuron Visualization}
\label{sec:example_neuron}

\begin{figure}
    \centering
    \begin{subfigure}{0.48\columnwidth}
        \includegraphics[width=\textwidth]{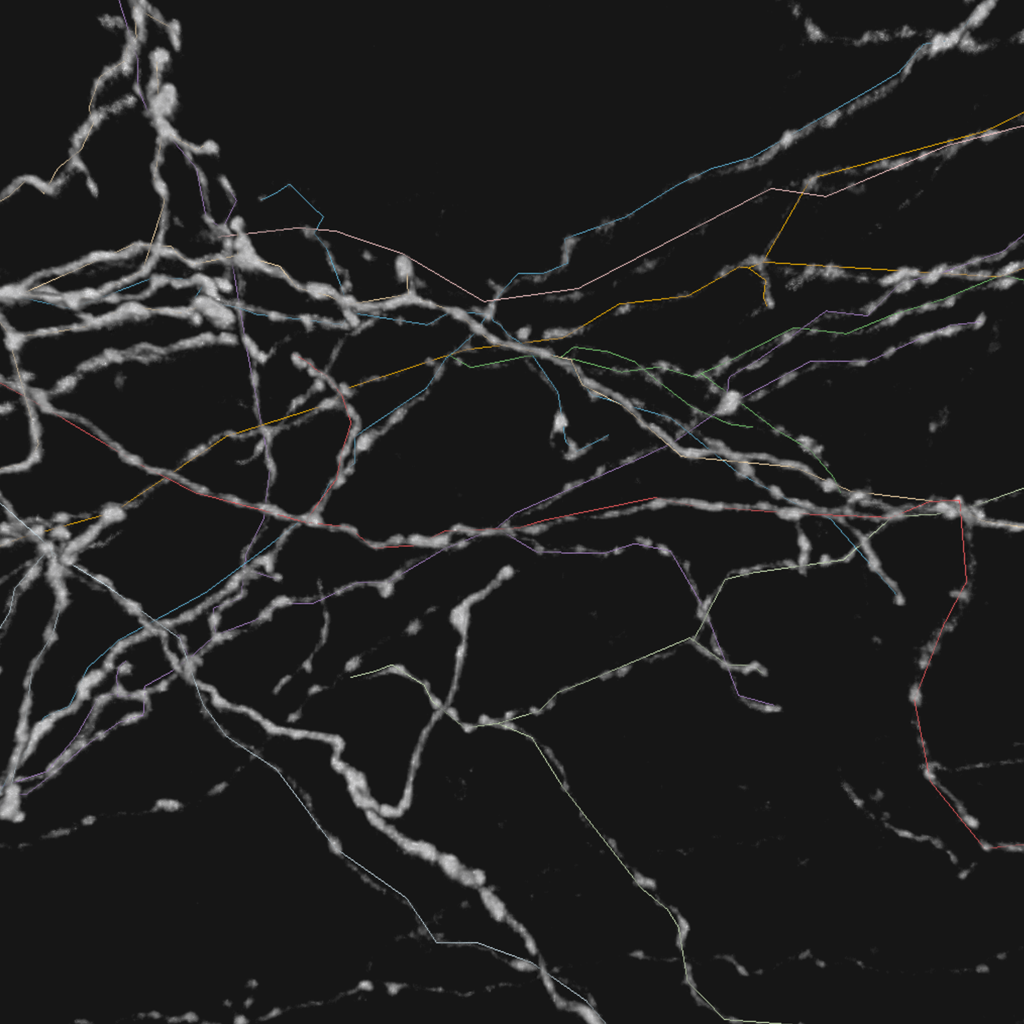}
    \end{subfigure}
    \begin{subfigure}{0.48\columnwidth}
        \centering
        \includegraphics[width=\textwidth]{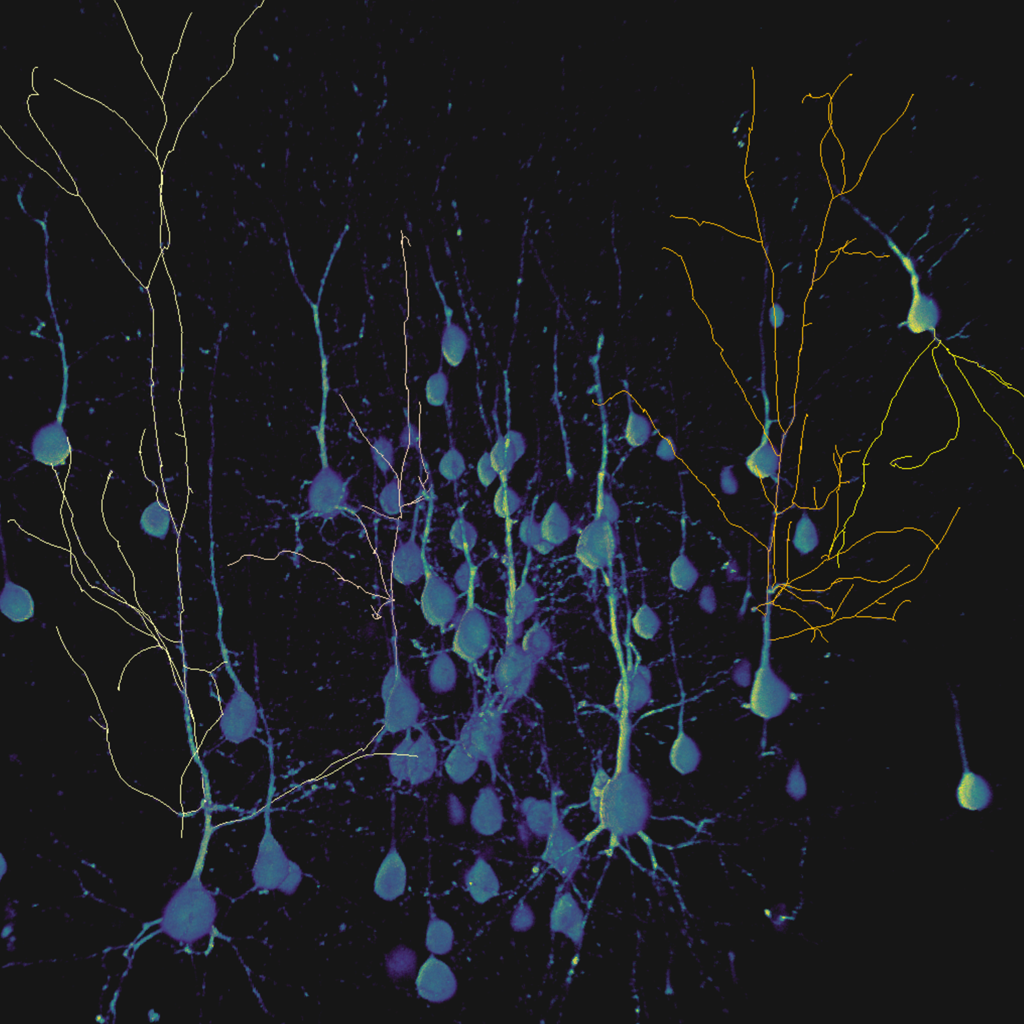}
    \end{subfigure}
    \vspace{-0.5em}
    \caption{\label{fig:neuron_vis}%
    Images of the neuron visualization application, used by
    neuronatomists to judge the quality of neuron reconstructions
    (rendered as lines) in the browser.\vspace{-1em}}
\end{figure}

\begin{table}
    \centering
    \relsize{-1}{
    \begin{tabular}{@{}lrrr@{}}
        \toprule
        Dataset & Native & Firefox 78 & Chrome 83 \\
        \midrule
        Layer 1 Axons (44MB, 76 files) & 0.87s & 1.8s & 2.6s \\
        Cell Bodies (263Mb, 314 files) & 4.1s & 8.6s & 16s \\
        \bottomrule
    \end{tabular}
    }
    \vspace{-0.5em}
    \caption{\label{tab:tiff_load}%
    Load times for the connectomics data sets using the native build of libtiff
    and the WebAssembly module in Firefox and Chrome.\vspace{-2em}}
\end{table}

The first example is a web application for visualization of connectomics data (\Cref{fig:neuron_vis}).
The application uses libtiff compiled to WebAssembly to import TIFF stacks which
are volume rendered using WebGL2.
TIFF files are loaded by writing the file data into Emscripten's virtual
in memory filesystem, after which the file name can be passed to \texttt{TIFFOpen}
and read through Emscripten's implementation of the POSIX file API.
The application can also import SWC files to display
neuron reconstructions with the volume data.
This tool was written to facilitate
collaboration on another project, allowing visualization researchers
to easily collect feedback from neuroanatomists on reconstruction quality.
The neuroanatomist is able to simply open the web page, upload the data and traces,
and interactively compare multiple reconstructions.
We chose WebGL2 for the rendering component
of this tool to ease deployment, as WebGPU is not yet available
outside preview browser builds.




\paragraph{Performance}
We compare time to load two connectomics data sets using
our WebAssembly build of libtiff and the native version in~\Cref{tab:tiff_load}.
The benchmarks are run on a Surface Pro 7 laptop with an i5-1035G4 CPU and 8GB
of RAM.
We find that, similar to the results reported by Jangda et al.~\cite{jangda_not_2019},
WebAssembly is 2-4$\times$ slower than native code.
The browser's implementation of the WebAssembly engine is an important factor,
with Firefox outperforming Chrome by 2$\times$.



\subsubsection{LIDAR Visualization}
\label{sec:example_lidar}

\begin{figure}[t]
    \centering
    \begin{subfigure}{0.49\columnwidth}
        \centering
        \includegraphics[width=\textwidth]{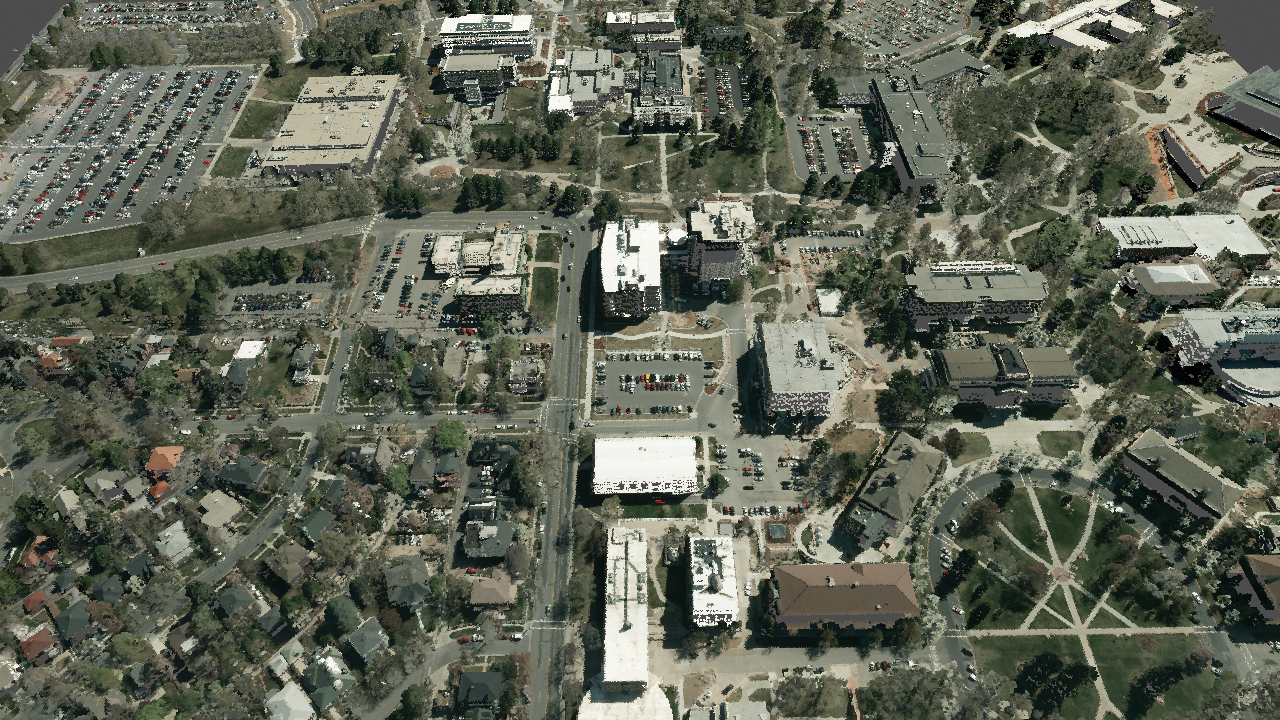}
        \caption{\label{fig:las_utah}%
        Utah 9M, points}
    \end{subfigure}
    \begin{subfigure}{0.49\columnwidth}
        \centering
        \includegraphics[width=\textwidth]{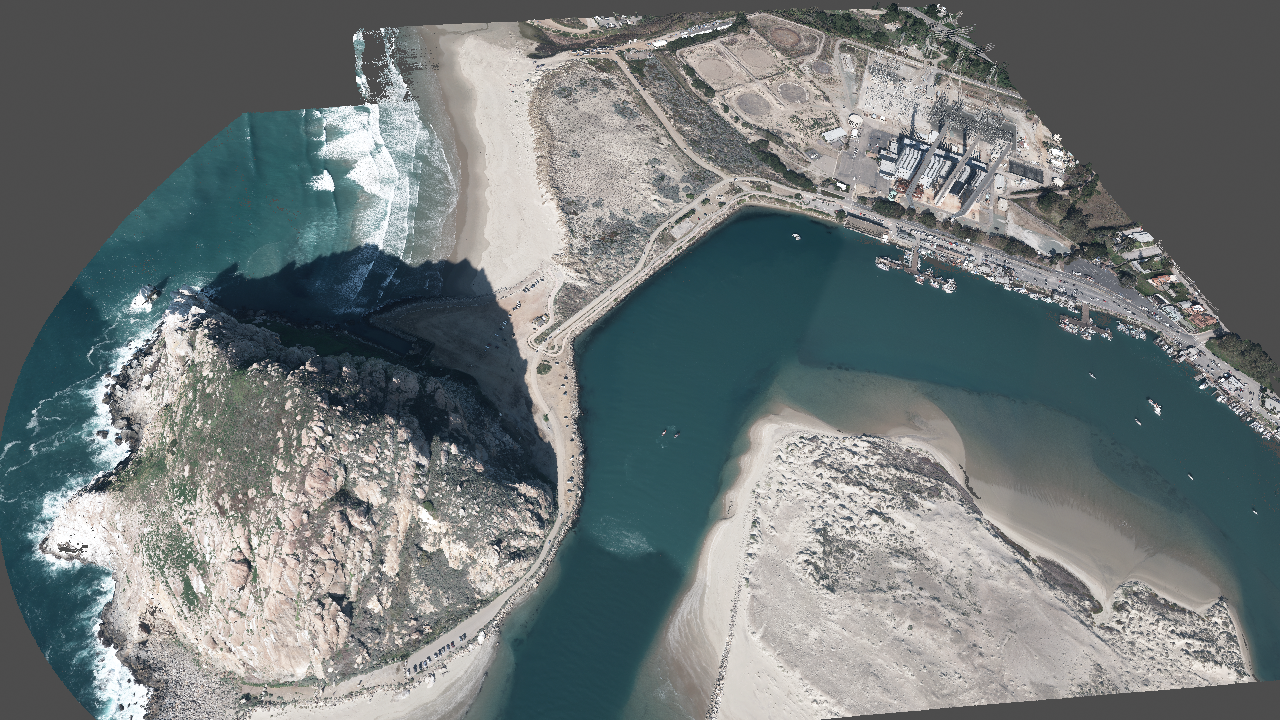}
        \caption{\label{fig:las_morro_40m}%
        Morro Bay, 40M points}
    \end{subfigure}
    \vspace{-0.75em}
    \caption{\label{fig:lidar_viewer}%
    Images from our WebGPU LIDAR renderer. At 1280$\times$720 the images render
    at (a) 100 FPS and (b) 35FPS on an RTX 2070, and (a) 14FPS and (b) 3.5FPS on
    a Surface Pro 7.\vspace{-1em}}
\end{figure}

Our second application example is LIDAR visualization, a common
task in geospatial applications.
To enable loading LIDAR data directly in the browser, we compiled the widely used library LAStools~\cite{lastools}
to WebAssembly. LAStools has no external dependencies, making
it easy to compile to Wasm; however, it is a C++ library and thus requires a C API
to be written or generated to be called from JavaScript. We wrote a minimal C API over
the library that supports loading files and retrieving the contained
points and colors to provide an illustrative
example of the process (see \Cref{lst:liblas_mem_sharing} and Github\footnotemark[3]).
Laz files uploaded by the user are written into Emscripten's virtual
filesystem to allow the library to open them.
The loaded points are then rendered as billboarded quads using WebGPU (\Cref{fig:lidar_viewer})
to display varying size round or square points.

\begin{table}[t]
    \centering
    \relsize{-1}{
    \begin{tabular}{@{}lrrr@{}}
        \toprule
        Dataset & Native & Firefox 80 & Chrome 86 \\
        \midrule
        Utah 9M & 3.8s & 9.2s & 29s \\
        Morro Bay 20M & 7.8s & 20s & 68s \\
        Utah 30M & 13s & 30s & 100s \\
        Morro Bay 40M & 16s & 41s & 133s \\
        \bottomrule
    \end{tabular}
    }
    \vspace{-0.5em}
    \caption{\label{tab:las_load_time}%
    Load times for the laz files using the native build of LASlib
    and the WebAssembly module in Firefox and Chrome.\vspace{-2em}}
\end{table}



\paragraph{Performance}
We compare the time to load various compressed LIDAR data files in laz format, from
9M to 40M points,
using the native and WebAssembly versions of LAStools in~\Cref{tab:las_load_time}.
For these benchmarks, we use Chrome Canary and Firefox Nightly, as they are
required for WebGPU support. The benchmarks are performed
on a Surface Pro 7. As before, we find that Firefox's WebAssembly engine
is 2-3$\times$ slower than native code, with Chrome's engine an additional
2-3.5$\times$ slower than Firefox. Although performance trails native code
overall, the loading times achieved for the data sets are likely acceptable in
an application.

In~\Cref{fig:lidar_viewer} we report the rendering performance achieved
by our LIDAR renderer on the smallest and largest data sets tested,
on both a Surface Pro 7 and a desktop with an RTX~2070 GPU.
We find that our basic renderer, which does not employ level of detail
or filtering to accelerate rendering, is still able to provide interactive rendering
of the largest data set even on the integrated GPU of the Surface Pro.

\subsubsection{Marching Cubes}
\label{sec:example_mc}
While our prior examples have focused on I/O and memory intensive tasks,
namely, loading compressed and complex data formats, another key
concern is the performance of compute intensive tasks.
We use Marching Cubes~\cite{lorenson_marching_1987} as a
representative proxy for the compute demands of common analysis tasks performed
in scientific visualization applications.
We evaluate the performance of serial implementations in C++, JavaScript,
and WebAssembly, and data-parallel variants in Vulkan and WebGPU.
Our data-parallel variants are similar to standard approaches using
CUDA~\cite{dias_cuda_based_2010,cuda_mc_example}:
a compute shader run over all voxels marks those containing the surface,
the active voxel IDs are then compacted using a GPU prefix sum and stream compaction.
For each active voxel, a compute shader determines the number
of output vertices and writes the result to a global buffer.
The offsets for each voxel and total number of vertices are computed using a GPU prefix sum.
A second pass over the active voxels computes and outputs the vertices to
produce a triangle soup.

\paragraph{Performance}
\label{sec:example_mc_perf}
\begin{figure}
    \centering
    \begin{subfigure}{0.49\columnwidth}
        \centering
        \includegraphics[width=\textwidth]{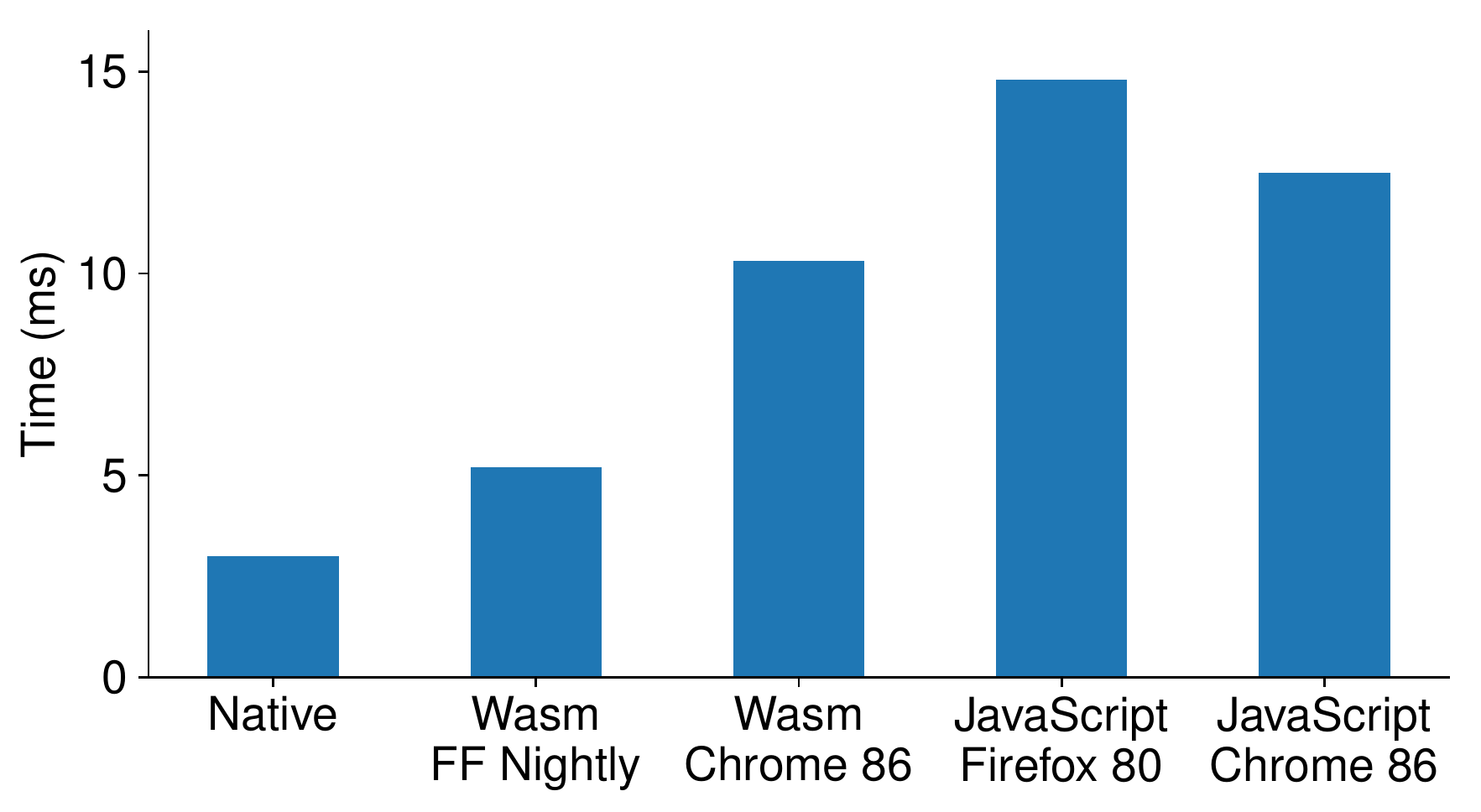}
        \caption{Fuel (64$^\text{3}$)}
    \end{subfigure}
    \begin{subfigure}{0.49\columnwidth}
        \centering
        \includegraphics[width=\textwidth]{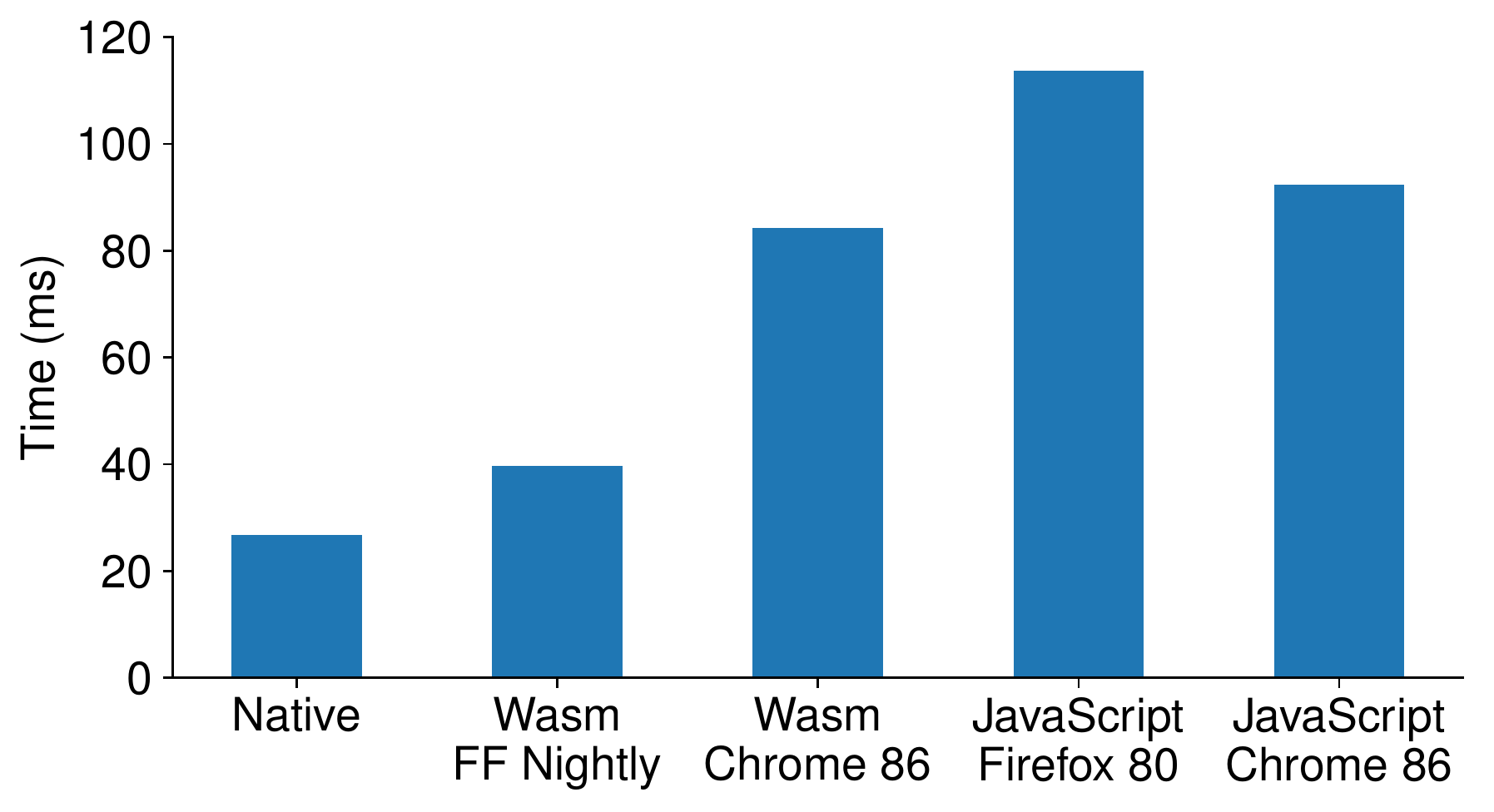}
        \caption{Hydrogen Atom (128$^\text{3}$)}
    \end{subfigure}
    \begin{subfigure}{0.49\columnwidth}
        \centering
        \includegraphics[width=\textwidth]{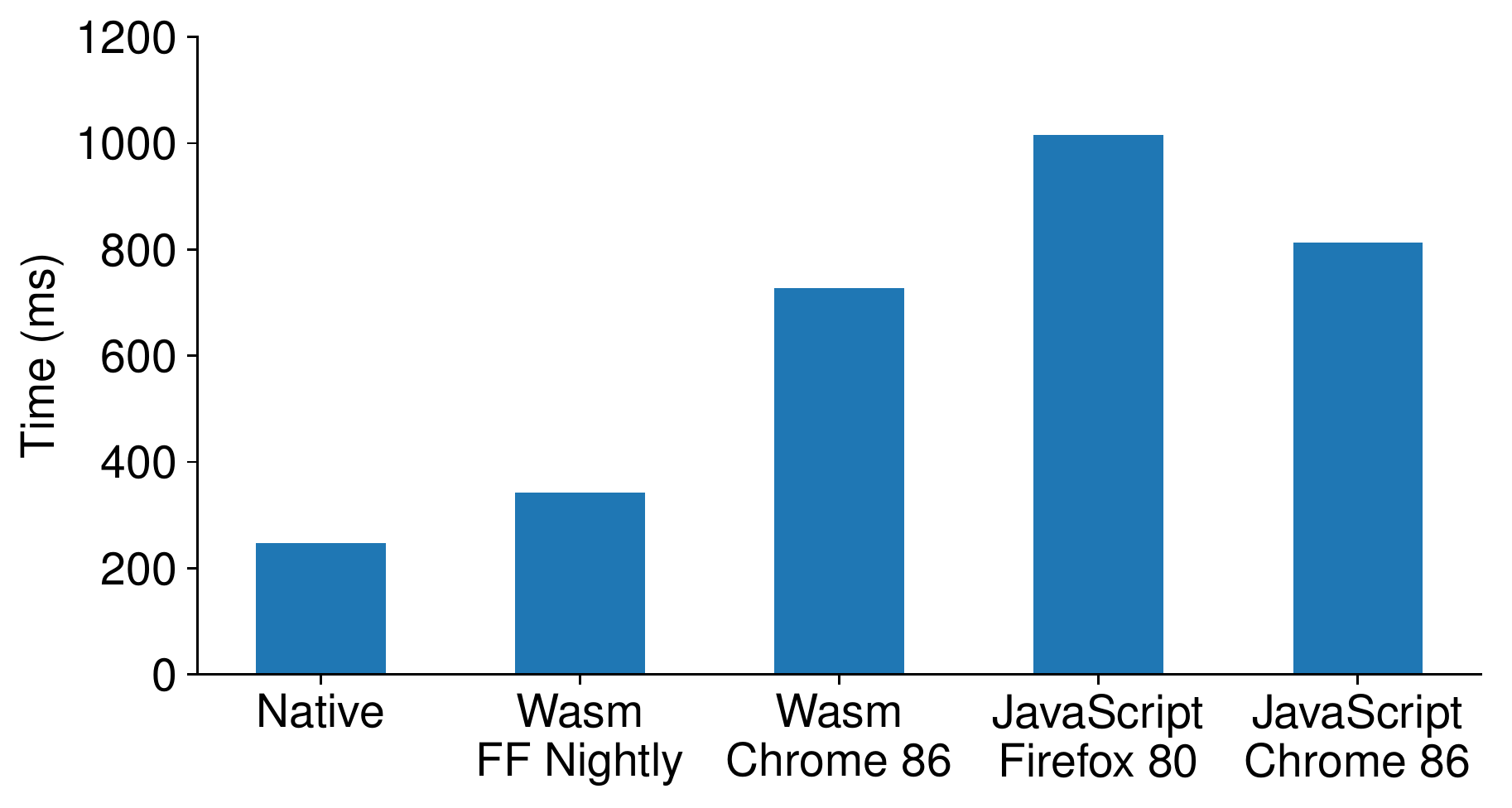}
        \caption{Bonsai (256$^\text{3}$)}
    \end{subfigure}
    \begin{subfigure}{0.49\columnwidth}
        \centering
        \includegraphics[width=\textwidth]{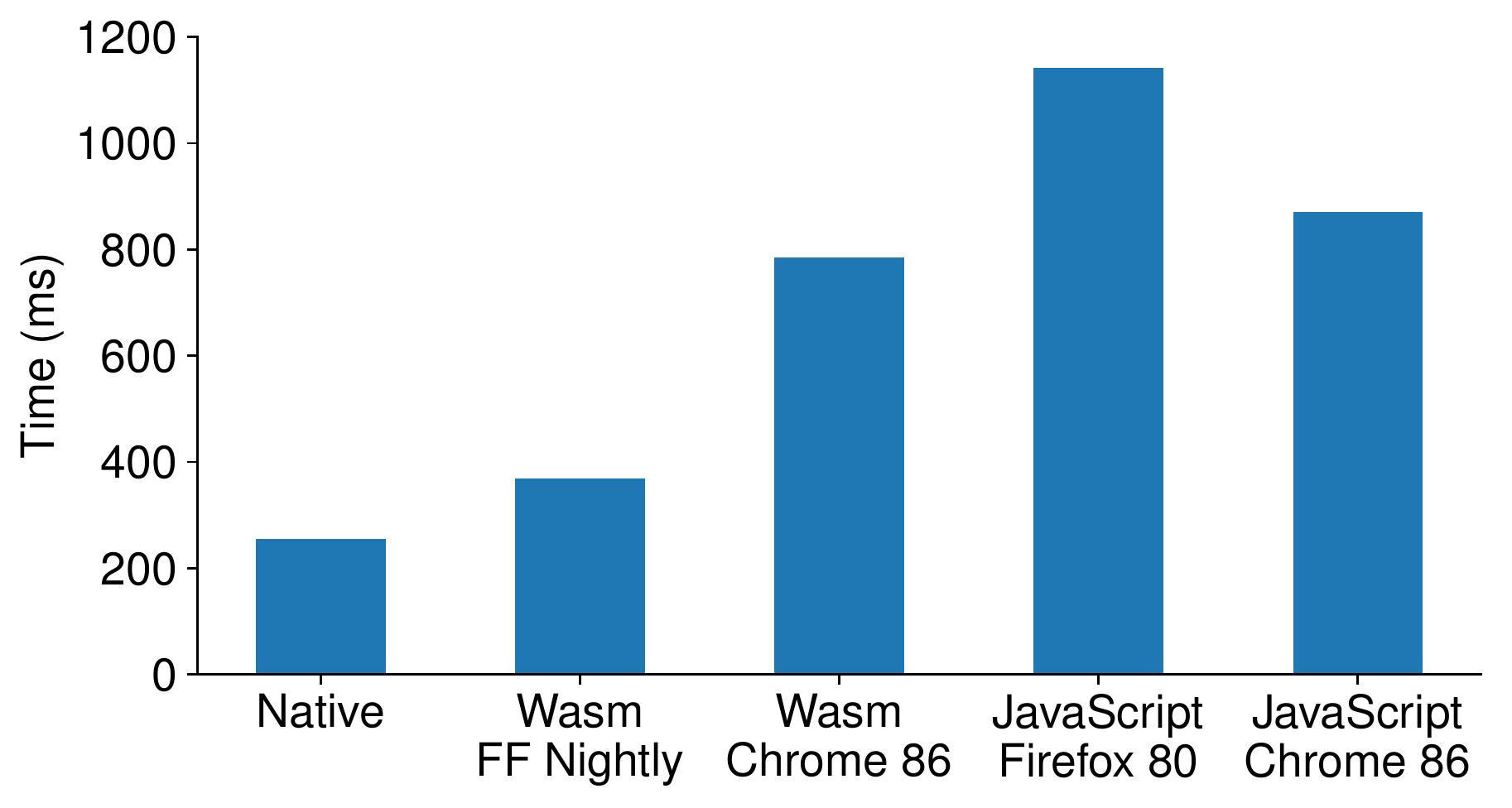}
        \caption{Skull (256$^\text{3}$)}
    \end{subfigure}
    \vspace{-0.75em}
    \caption{\label{fig:serial_mc_perf}%
    Performance of our serial C++, JavaScript, and WebAssembly versions
    of Marching Cubes on 100 random isovalues.\vspace{-1em}}
\end{figure}

We compare our serial implementations on the Surface Pro 7 on
four small data sets, ranging in size from $64^3$ to $256^3$.
As before, we use Chrome Canary and Firefox Nightly.
We compute 100 random isovalues using each variant and report the
average time to compute the surface.
The 100 random isovalues are sampled over a range covering surfaces
typically of interest to users, excluding noise and values so high as
to produce few output triangles.

For our serial implementations (\Cref{fig:serial_mc_perf}),
we find that our WebAssembly version run in Firefox
is on average only 1.5$\times$ slower than native code.
Firefox's JavaScript engine is on average 4.4$\times$ slower than native code,
with the performance gap between JavaScript and WebAssembly increasing with data set size.
Chrome's WebAssembly performance is on par with its JavaScript performance,
averaging 3.2$\times$ and 3.6$\times$ slower than native code, respectively.

We compare our data-parallel versions on the Surface Pro 7 and a
desktop with an RTX~2070 (\Cref{tab:mc_naive_parallel_perf}),
and test on an additional $512^3$ data set.
In each benchmark, we discard the first computation as we have found WebGPU to
have a high first launch overhead.
On the Surface Pro, we find moderate improvement over the fastest serial implementation
in the browser on larger data sets, although both the Vulkan and WebGPU variants achieve performance
just on par with the serial C++ version. When compared to the fastest WebAssembly
version, the Bonsai and Skull see performance improvements of $\sim$1.3$\times$.
On the RTX~2070, we find significant performance improvements over the serial implementations,
achieving a 7$\times$ improvement on the $256^3$ data sets
over WebAssembly in Firefox, and a 5$\times$ improvement over our serial native implementation.

The most exciting takeaway from our evaluation is that
the performance of WebGPU is typically on par with native Vulkan
code, indicating that little performance is lost when
moving GPU compute kernels to the browser. On smaller data sets, we
find WebGPU to be slower, where overhead in the WebGPU mapping layer
to the underlying native graphics API may impact performance.
This overhead is hidden on larger data sets, where the computation
dominates the total execution time.


\begin{table}
    \centering
    \begin{subtable}{0.49\columnwidth}
        \centering
        \relsize{-1}{
        \begin{tabular}{@{}lrr@{}}
            \toprule
            Dataset & Vulkan & WebGPU \\
            \midrule
            Fuel $64^3$ & 25ms & 26ms \\
            Hydr. $128^3$ & 58ms & 86ms \\
            Bonsai $256^3$ & 228ms & 261ms \\
            Skull $256^3$ & 279ms & 282ms \\
            Plasma $512^3$ & 1357ms & $*$ \\
            \bottomrule
        \end{tabular}
        }
        \caption{\label{tab:mc_naive_parallel_perf_surface}%
        Surface Pro 7}
    \end{subtable}%
    \begin{subtable}{0.49\columnwidth}
        \centering
        \relsize{-1}{
        \begin{tabular}{@{}lrr@{}}
            \toprule
            Dataset & Vulkan & WebGPU \\
            \midrule
            Fuel $64^3$ & 2ms & 18ms \\
            Hydr. $128^3$ & 8ms & 18ms \\
            Bonsai $256^3$ & 49ms & 49ms \\
            Skull $256^3$ & 70ms & 50ms \\
            Plasma $512^3$ & 322ms & 329ms \\
            \bottomrule
        \end{tabular}
        }
        \caption{\label{tab:mc_naive_parallel_perf_rtx2070}%
        RTX 2070}
    \end{subtable}
    \vspace{-0.5em}
    \caption{\label{tab:mc_naive_parallel_perf}%
    Performance of our naive data-parallel Marching Cubes implementation
    on 100 random isovalues. We find that WebGPU performance is typically
    on par with native Vulkan code. $*$ failed on the Surface Pro.\vspace{-2em}}
\end{table}

\subsubsection{ZFP Decompression}
\label{sec:example_zfp}
When visualizing large-scale data on a remote client (e.g., the 1TB volume shown in~\Cref{fig:teaser}),
transferring the entire data set to the client can be impractical or impossible.
To reduce bandwidth and memory requirements, we can transfer a compressed version of the data,
that the client can decompress as needed.
ZFP~\cite{lindstrom_fixed_rate_2014} provides fast and high-quality
compression of floating point scientific data, and is especially suited to large volumetric data.
ZFP compresses volumetric data in $4^3$ blocks, using fixed or variable bitrate compression.
ZFP's fixed rate compression mode can be quickly compressed and decompressed in parallel on the GPU,
and a CUDA implementation is provided with the library. To enable fast decompression of
large data in the browser, we ported ZFP's CUDA decompressor to WebGPU, and to Vulkan
to provide a native comparison point.


\paragraph{Performance}

\begin{table}
    \centering
    \begin{subtable}{0.99\columnwidth}
        \centering
        \relsize{-1}{
        \begin{tabular}{@{}lrrr@{}}
            \toprule
            Dataset & CPU (Serial) & Vulkan & WebGPU \\
            \midrule
            Skull $256^3$ & 329MB/s & 756MB/s & 697MB/s \\
            Plasma $512^3$ & 349MB/s & 898MB/s & 777MB/s \\
            \bottomrule
        \end{tabular}
        }
        \vspace{-0.5em}
        \caption{\label{tab:zfp_surface_decompress}%
        Surface Pro 7\vspace{0.25em}}
    \end{subtable}
    \begin{subtable}{0.99\columnwidth}
        \centering
        \relsize{-1}{
        \begin{tabular}{@{}lrrr@{}}
            \toprule
            Dataset & CUDA & Vulkan & WebGPU \\
            \midrule
            Skull $256^3$ & 67.1GB/s & 14.8GB/s & 7.6GB/s \\
            Plasma $512^3$ & 67.8GB/s & 19.4GB/s & 11.8GB/s \\
            Miranda $1024^3$ & 113GB/s & 19.6GB/s & 13.7GB/s \\
            \bottomrule
        \end{tabular}
        }
        \vspace{-0.5em}
        \caption{\label{tab:zfp_rtx2070_decompress}%
        RTX 2070}
    \end{subtable}
    \vspace{-0.75em}
    \caption{\label{tab:zfp_decompress}%
    ZFP decompression performance of our WebGPU implementation compared
    to a native Vulkan version and ZFP's original CUDA and serial decompressor.
    We find that neither our Vulkan nor WebGPU implementation is on par
    with the original CUDA implementation, though are still capable of
    fast parallel decompression.
    \vspace{-2.5em}}
\end{table}

We compare our WebGPU and Vulkan ports of ZFP's CUDA decompressor on three data
sets, on both a Surface Pro 7 and RTX~2070 (\Cref{tab:zfp_decompress}).
For each data set, we compress it at three different bitrates, two, four, and eight,
and report the average decompression performance of 10 runs.
We use ZFP prerelease version 0.5.5-rc1 built from Github.
As CUDA is not available on the Surface, and parallel decompression with OpenMP
is not supported in ZFP, we compare our WebGPU and Vulkan variants to ZFP's serial
decompressor on the CPU. The decompressed output of the largest data set,
Miranda (4GB), does not fit in the Surface's 3.8GB VRAM, and thus we
evaluate only the smaller data sets.

Although we achieve slightly over $2\times$ faster decompression on the Surface
compared to the serial CPU decompressor, both our WebGPU and Vulkan versions
are slower than the CUDA decompressor, by up to $8\times$ and $6\times$ on the Miranda.
We also find the WebGPU implementation to trail the Vulkan version slightly,
as observed previously. 
Although our implementations have room for improvement
compared to ZFP's CUDA decompressor,
the performance improvements over serial decompression are substantial.


\section{Parallel Isosurface Extraction from Block-Compressed Data using WebGPU}

\begin{figure*}
    \centering
    \vspace{-1em}
    \includegraphics[width=0.95\textwidth]{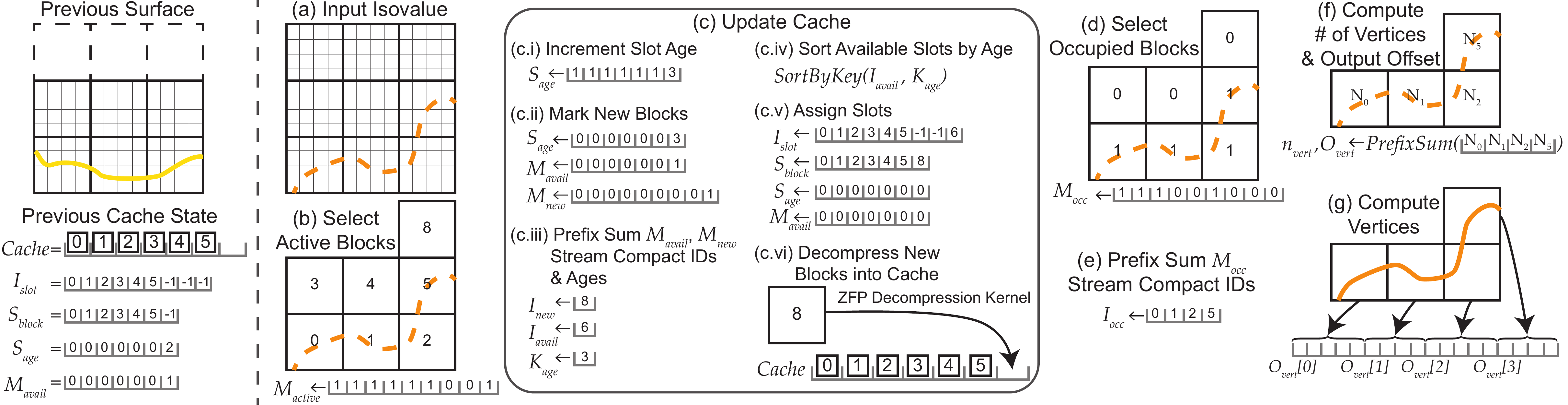}
    \vspace{-0.75em}
    \caption{\label{fig:algorithm_sketch}%
    An illustration of our parallel isosurface algorithm in 2D.
    Starting from an existing cache used to compute a previous surface (or an empty one)
    we wish to compute the surface at the new isovalue in (a).
    (b) We find the blocks that contain the isovalue
    or are needed to provide neighbor data for blocks containing it.
    (c) We determine which of these blocks are already in the cache
    and those that must be decompressed and added to it (also see \Cref{alg:cache_update}).
    (d) The active blocks are filtered down to just those that will output vertices,
    (e) after which we compute the number of vertices that will be output by each
    block and (f) output them to a single vertex buffer.
    \vspace{-2em}}
\end{figure*}


Based on the performance results observed on the example
applications, we conclude that the browser can provide a capable environment for
scientific visualization tasks.
However, working with large data sets in the browser using WebGPU is challenging
due to limitations on network transfer bandwidth and GPU memory,
requiring a combination of adaptive precision and resolution
compression (e.g.,~\cite{lindstrom_fixed_rate_2014,hoang-precision-tvcg-2019,hoang_efficient_2021}).

To enable interactive isosurface computation on large-scale data
in the browser, we build on the following observations.
First, the isosurfaces of interest to users typically occupy a sparse subset
of the volume, which likely does fit in memory.
Users are also likely to explore nearby isosurfaces, touching the same
regions of the volume repeatedly.
Third, ZFP's fixed-rate compression mode allows specific $4^3$ blocks to be decompressed,
without decompressing the entire data set.
Finally, we find WebGPU performance to be on par with native, and thus pushing
as much computation as possible to the GPU will be beneficial for
processing large data sets.

These observations motivate the design of our GPU parallel algorithm
for isosurface computation on block-compressed data (\Cref{fig:algorithm_sketch}
and on GitHub\footnote{\href{https://github.com/Twinklebear/webgpu-bcmc}{https://github.com/Twinklebear/webgpu-bcmc}}).
We begin by uploading the ZFP fixed-rate compressed data to the GPU,
allowing us to decompress blocks as needed without additional interaction with the CPU.
To compute the isosurface, we first find the blocks containing the data
required to compute it (\Cref{sec:mc_select_active_blocks}).
We then determine which of these blocks are in the cache or must be added to it (\Cref{sec:mc_lru_cache}).
Each new block is assigned a slot in the cache and decompressed into it (\Cref{sec:mc_decompression}).
Active blocks are processed by loading them and the
required neighbor voxels into shared memory (\Cref{sec:mc_load_shmem}).
For each active block, we compute the number of vertices output
by its voxels (\Cref{sec:mc_compute_output_locations}).
Finally, for each block that will output vertices, we compute
the vertices to output a triangle soup (\Cref{sec:mc_vertex_compute}).



\subsection{Selecting Active Blocks}
\label{sec:mc_select_active_blocks}

\begin{figure}
    \centering
    \includegraphics[width=0.4\columnwidth]{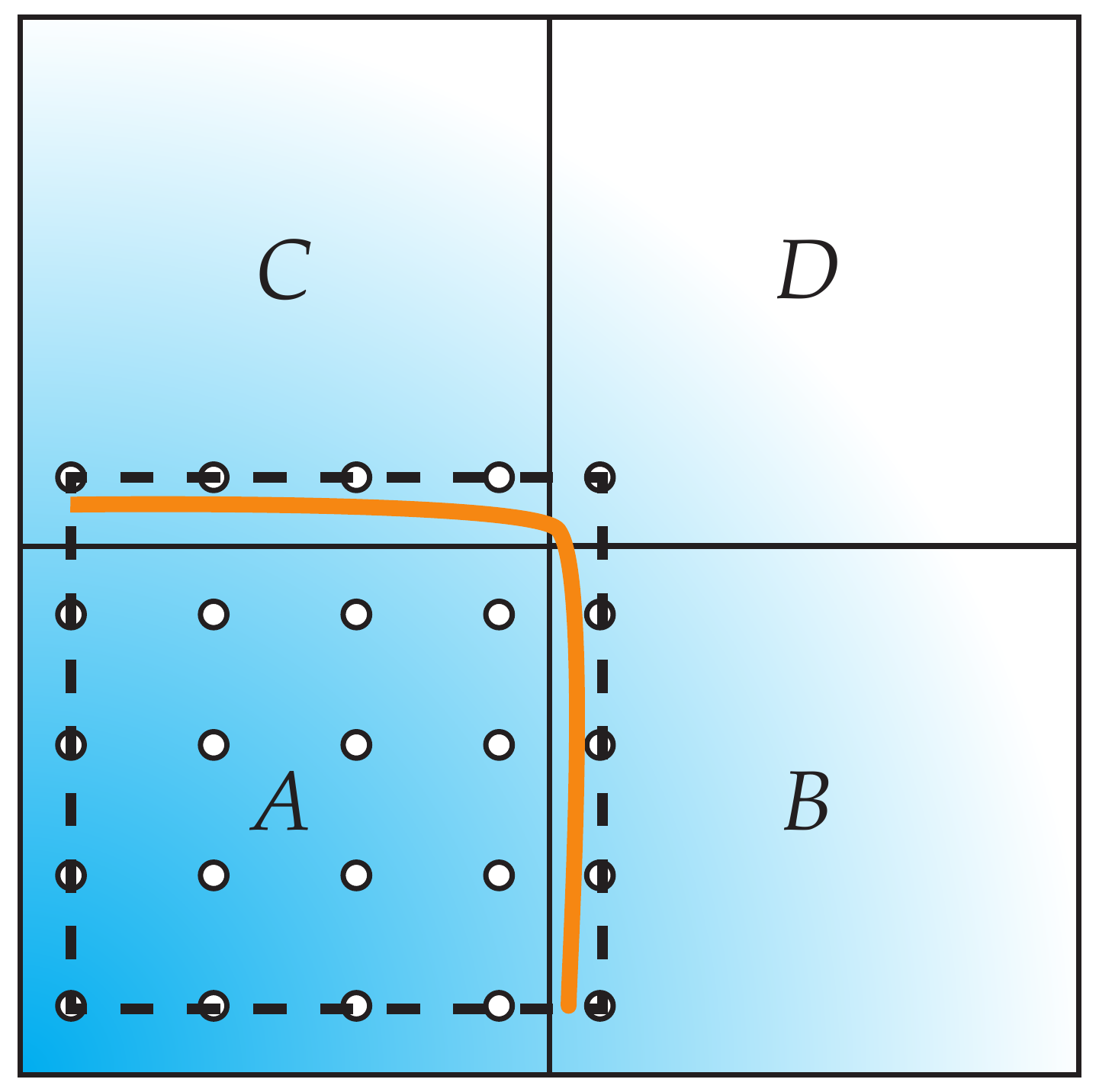}
    \vspace{-1em}
    \caption{\label{fig:block_dual_grid}%
    The dual grid of block $A$ (dashed) overlaps its neighbors $B$, $C$, $D$,
    who must also be decompressed to provide the data required to compute
    the surface (orange).
    \vspace{-2em}}
\end{figure}

As done by Liu et al.~\cite{liu_parallel_2016}, we precompute and store
the value range of each block when loading the data. To find active
blocks, we run a compute shader over the blocks that marks a block as
active if its range contains the isovalue (\Cref{fig:algorithm_sketch}b).
However, in contrast to Liu et al., we do not
have access to the full volume in a 3D texture.
Thus, marking only the blocks containing the isovalue as active is not sufficient,
since the data required
for their dual grid cells sharing vertices with neighboring blocks would be
missing (\Cref{fig:block_dual_grid}).
To ensure the neighbor data is available, we mark
a block as active if its range or the union of its range
and any of its neighbors' ranges contain the isovalue.
The output of our kernel is the active mask list $M_{active}$.
The list of active block IDs, $I_{active}$, is computed using a GPU prefix sum
and specialized stream compaction, ``StreamCompactIDs''.
The specialized compaction writes the element's index in the active mask list at the output
offset (computed in the prefix sum) instead of the element's value, compacting the IDs of the active elements.

\subsection{GPU-driven LRU Block Cache}
\label{sec:mc_lru_cache}

Given $M_{active}$ and $I_{active}$,
we determine which active blocks are already in the cache
and which must be decompressed and added to it.
This task is performed using an entirely GPU-driven LRU cache,
allowing us to avoid reading back $M_{active}$ or $I_{active}$ to the host
and to minimize the amount of serial computation.

The cache is comprised of a growable number of ``slots'', within which a decompressed
block can be stored.
The initial number of slots is set by the user, e.g., enough to store 10\% of the blocks in the volume.
We use two arrays to track the status of each slot: $S_{age}$, which stores
the age of the item in the slot, and $S_{block}$, which stores the ID of the block in the slot
or -1 if the slot is empty.
An additional array, $I_{slot}$, stores the ID of the slot occupied by each block, or -1 if
the block is not cached.
Finally, we maintain a buffer storing the actual data for each slot, containing the decompressed $4^3$ block.


\begin{algorithm}[t]
    \vspace{0.25em}
    \begin{algorithmic}[1]
        \Function{UpdateCache}{$M_{active}$}
            \State $IncrementSlotAge(S_{age})$
            \State $M_{new}, M_{avail} \gets MarkNewBlocks(M_{active}, S_{age}, I_{slot})$
            \State $n_{new}, O_{new} \gets PrefixSum(M_{new})$ \Comment{Exit if $n_{new} = 0$}
            \State $n_{avail}, O_{avail} \gets PrefixSum(M_{avail})$ \Comment{Grow if $n_{new} > n_{avail}$}
            \State $I_{new} \gets StreamCompactIDs(M_{new}, O_{new})$
            \State $I_{avail} \gets StreamCompactIDs(M_{avail}, O_{avail})$
            \State $K_{age} \gets StreamCompact(M_{avail}, O_{avail}, S_{age})$
            \State $SortByKey(I_{avail}, K_{age})$ \Comment{Sort available slots by age}
            \State $AssignSlots(I_{new}, I_{avail}, S_{age}, S_{block}, I_{slot})$
            \State $DecompressBlocks(I_{new}, I_{slot})$
        \EndFunction
    \end{algorithmic}
    \vspace{-1em}
    \caption{\label{alg:cache_update}%
    Our GPU-driven cache update algorithm. Each function call corresponds
    to a compute shader dispatch or data-parallel primitive (prefix sum, stream compact, sort)
    run on the GPU.
    \vspace{-2em}}
\end{algorithm}

Our GPU parallel cache update proceeds as shown in~\Cref{alg:cache_update}
(also see~\Cref{fig:algorithm_sketch}c).
First, we increment each slot's age in parallel using a compute shader.
We then compute a list $M_{new}$ that marks, for each block, if it must
be newly decompressed and added to the cache; and update the list $M_{avail}$
that marks, for each slot, if it is available.
To compute $M_{new}$ and $M_{avail}$, we run a compute shader for each block $b$ that checks
if it is active (i.e., $M_{active}[b] = 1$).
If the block is active and cached (i.e., $I_{slot}[b] \neq -1$), the thread resets
the age of its slot, marks the slot unavailable in $M_{avail}$, and marks the block as not
new in $M_{new}$. If the block is active and not cached, it is marked as new.
If a block is cached and not active its slot is marked as available, making it
a candidate for eviction.
We then perform a GPU prefix sum on $M_{new}$ and $M_{avail}$
to compute the number of new blocks and the number of available slots.
If there are no new blocks, we can terminate the update,
if fewer slots are available than are needed, the cache is grown.
New slots added when growing the cache are marked available and assigned a high age.

We then use ``StreamCompactIDs'' to compute the list of new block IDs, $I_{new}$, and available
slots, $I_{avail}$. The active list and output offsets come from $M_{new}$, $O_{new}$ and $M_{avail}$, $O_{avail}$,
respectively.
The available slot ages, $K_{age}$, are computed with a standard stream compaction
using the output offsets $O_{avail}$.
$I_{avail}$ is then sorted in descending order by age using a GPU sort by key,
where $K_{age}$ contains the keys.
Finally, we assign each new block a slot $s$ from $I_{avail}$ using a compute
shader run for each new block.
If a block $p$ was previously stored in the slot (i.e., $S_{block}[s] \neq -1$),
it is evicted by setting $I_{slot}[p] = -1$.
The new block is then set as the item in the slot,
the slot assigned to the block in $I_{slot}$,
the slot marked unavailable, and the slot age reset to 0.


\subsubsection{Decompression of Blocks into Cache Slots}
\label{sec:mc_decompression}


We use our WebGPU port of ZFP's CUDA decompressor
from \Cref{sec:example_zfp}, modified to decompress just the blocks specified
in $I_{new}$. Each new block $b$ is decompressed by a thread on the GPU in a compute shader.
The thread writes the $4^3$ decompressed block as a row-major array into the cache
slot assigned to the block ($I_{slot}[b]$).
This approach has the added benefit of eliminating the scattered writes required
when outputting the entire volume, improving the write access patterns
of each decompression thread.

\subsection{Loading Blocks and Neighbor Voxels into Shared Memory}
\label{sec:mc_load_shmem}
To compute the vertices of each block, we need the data for it and its neighbors
in the $+x/y/z$ direction, with which the block's dual grid shares vertices (\Cref{fig:block_dual_grid}).
As we do not store the full volume, accessing the neighbor voxels cannot be
done with a 3D texture lookup as done by Liu et al.~\cite{liu_parallel_2016}.
Instead, we load the data for the block's dual grid into shared memory to
enable fast access to the data by the thread group processing the block.
This step is performed by both the vertex output location computation (\Cref{sec:mc_compute_output_locations})
and the final vertex computation (\Cref{sec:mc_vertex_compute}).
In both steps, we assume the block has neighbors along all axes,
i.e., a $4^3$ dual grid, and we run a thread group per block with 64 threads,
one thread per dual cell.

First, the group loads the 64 voxels of the $4^3$ block into shared memory, with each thread
loading a voxel.
If the block has neighbors to its positive side
we increase the block dimensions along the corresponding axis by one.
However, we cannot assume the required
face, edge, and corner neighbors exist, as only the active blocks are guaranteed to be
decompressed and in the cache. For example, a block that is marked active because it
is needed by a neighbor may not have any of its $+x/y/z$ neighbors active, and would
thus attempt to load invalid data.
After tentatively increasing the block's dimensions based on its location in the grid,
we test if the required neighbors are active in $M_{active}$.
If the required face, edge, or corner neighbor in some direction is not active,
we reset the block's size along that direction to 4.

We then load the required neighbor data
from the neighbor blocks. The threads responsible for the block's face,
edge, and corner voxels are also responsible for loading the neighboring face, edge,
and corner voxels, respectively.
Finally, we synchronize the thread group using a barrier to wait until the
data is loaded.

%
%

\subsection{Computing Vertex Output Locations}
\label{sec:mc_compute_output_locations}
Our next step is to determine the output offsets for each block's vertices
and the total number of vertices to be output.
Before doing so, we run a compute shader over the active blocks to
determine the subset that will output vertices (\Cref{fig:algorithm_sketch}d).
Each thread group loads a block into shared memory as described in~\Cref{sec:mc_load_shmem}. 
Each thread then checks if its dual cell will output a triangle and writes this result to shared memory.
If the block will output a triangle, thread 0 marks the block
as occupied in $M_{occ}$; otherwise, it is marked as empty.
The list of occupied block IDs, $I_{occ}$,
are found through a GPU prefix sum over $M_{occ}$ and ``StreamCompactIDs'' (\Cref{fig:algorithm_sketch}e).
It is not sufficient to mark unoccupied each block that was marked
active as a result of the union of it and one of its neighbors' ranges, as the
block's dual cells shared with the neighbor may output vertices that
the block is responsible for computing (\Cref{fig:block_dual_grid}).
However, it may be possible to determine
this by checking which side the neighbor was on.



We then compute the output offsets for each occupied block's vertices (\Cref{fig:algorithm_sketch}f).
After loading the block and neighbor data into shared memory, we proceed
as described by Liu et al.~\cite{liu_parallel_2016} to compute the vertex output offsets.
Each thread computes the number of vertices that will be output by its dual
cell and writes this value to shared memory. The thread group then performs a sum reduction
to compute the total number of vertices that will be output
by the block, which thread 0 writes to global memory.
The per block output offsets and total output size are computed by
a GPU prefix sum over the per block vertex counts.
As suggested by Liu et al., we do not perform
a global prefix sum or store the per voxel prefix sums computed within each block.

\def\tableimgwidth{0.2\columnwidth}
\begin{table*}
    \vspace{-1.5em}
    \centering
    \relsize{-1}{
    \begin{tabular}{@{}m{\tableimgwidth}ccrrrrrr@{}}
        \toprule
        Image & Original & \makecell{Adaptive Precision \\ \& Resolution} & Isovalue & \% Active & Cache Size & \% Occupied & Triangles & Vertex Data \\
        \midrule
        \includegraphics[width=\tableimgwidth]{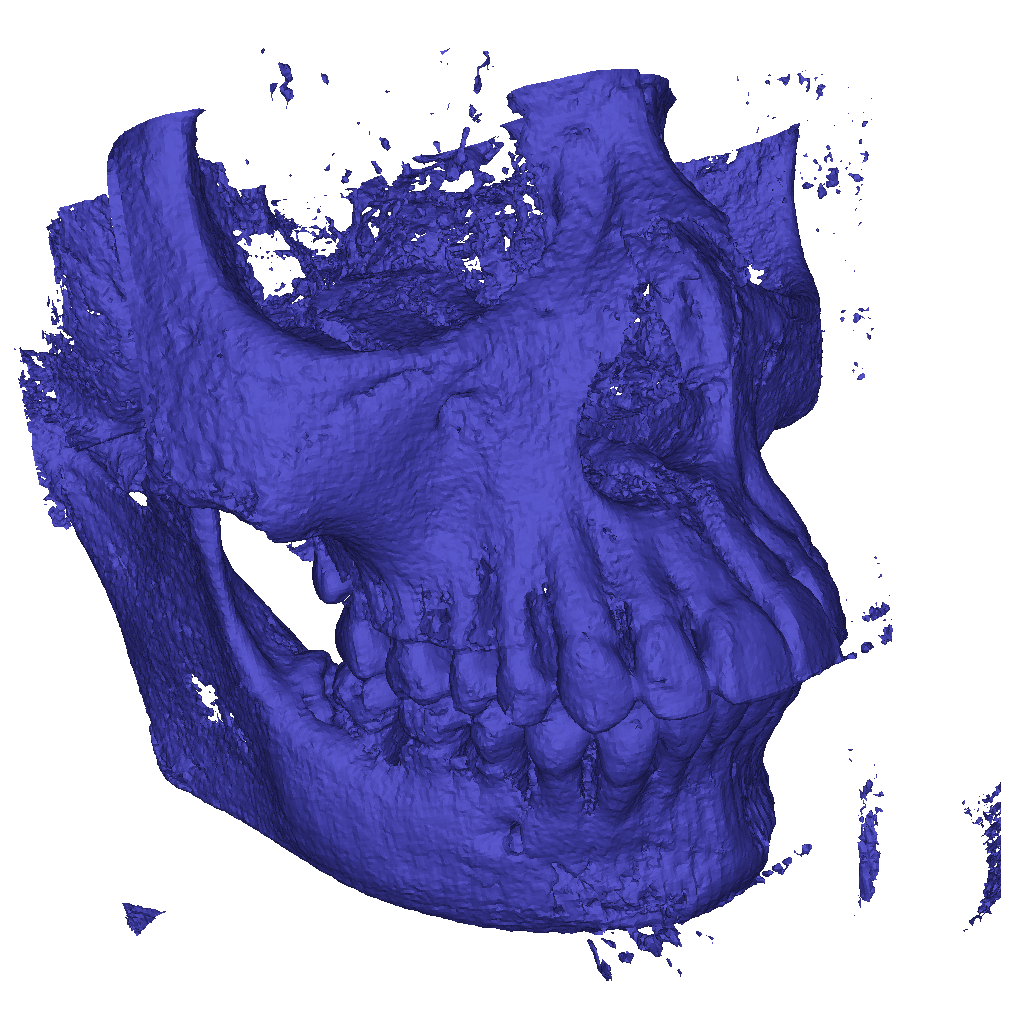}
            & \makecell{256$^\text{3}$ \\ 67.1MB}
            & \makecell{4.1MB \\ $\approx\frac{1}{16}$ data size}
            & 39 & 38.9\% & 26MB & 19.8\% & 2.1M & 51MB \\

        \includegraphics[width=\tableimgwidth]{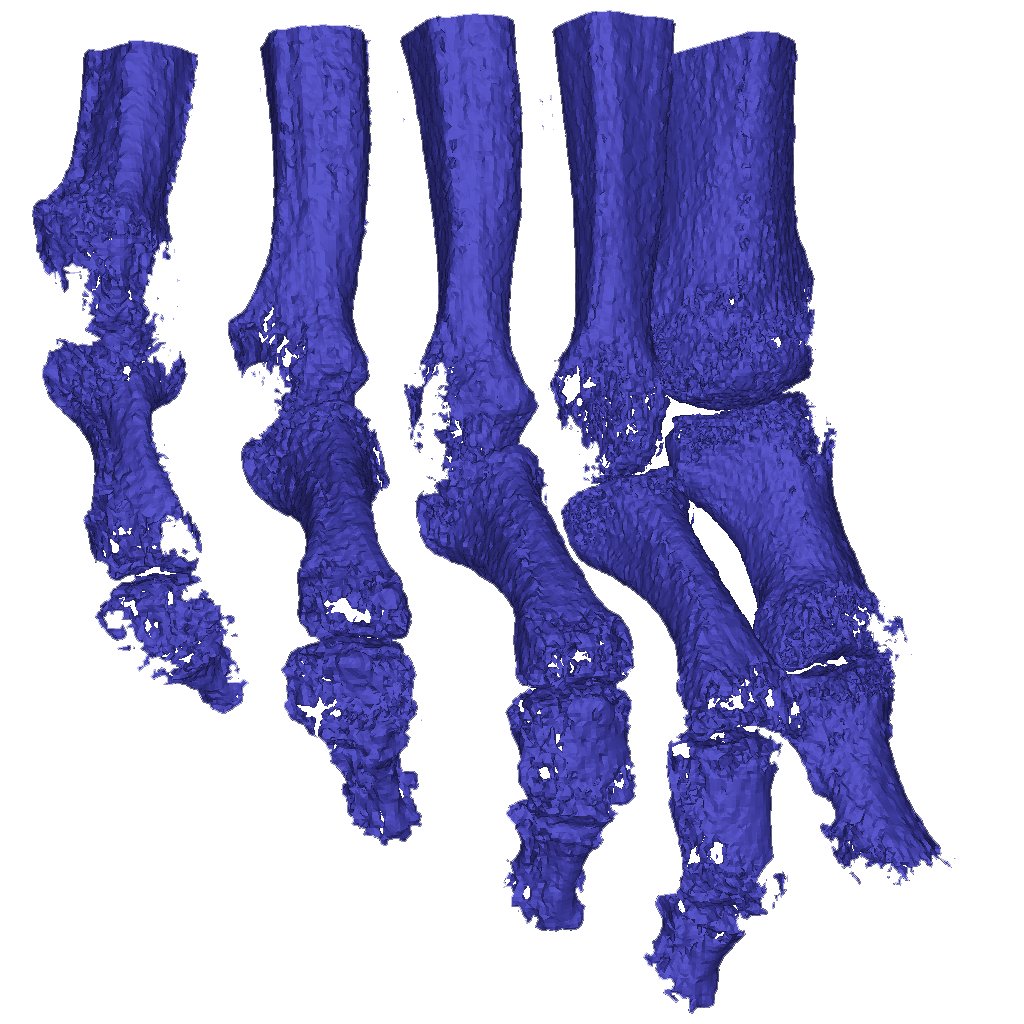}
            & \makecell{256$^\text{3}$ \\ 67.1MB}
            & \makecell{4.1MB \\ $\approx\frac{1}{16}$ data size}
            & 117 & 11.7\% & 7.8MB & 5.9\% & 743K & 18MB \\

        \includegraphics[width=\tableimgwidth]{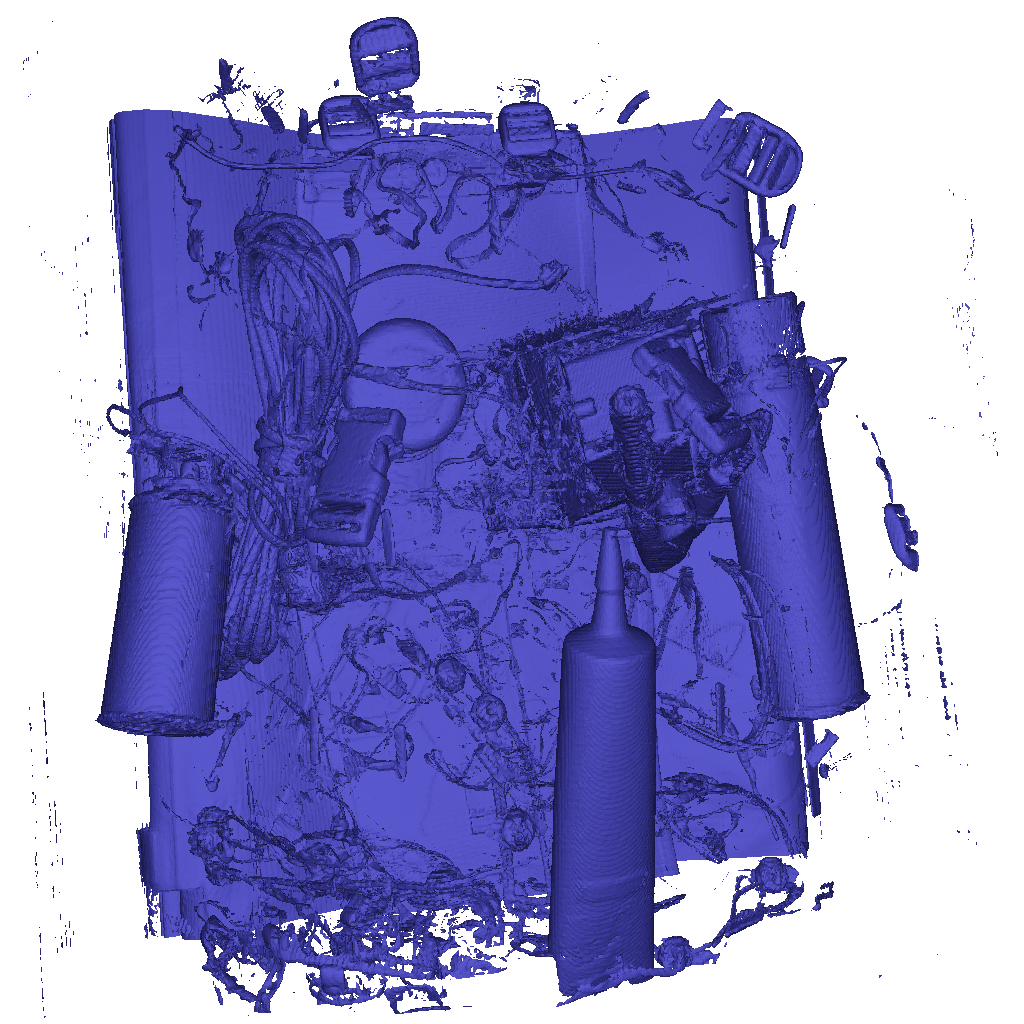}
            & \makecell{512$\times$512$\times$373 \\ 391MB}
            & \makecell{49.3MB \\ $\approx\frac{1}{8}$ data size}
            & 409 & 22.6\% & 89MB & 9.9\% & 6.2M & 148MB \\

        \includegraphics[width=\tableimgwidth]{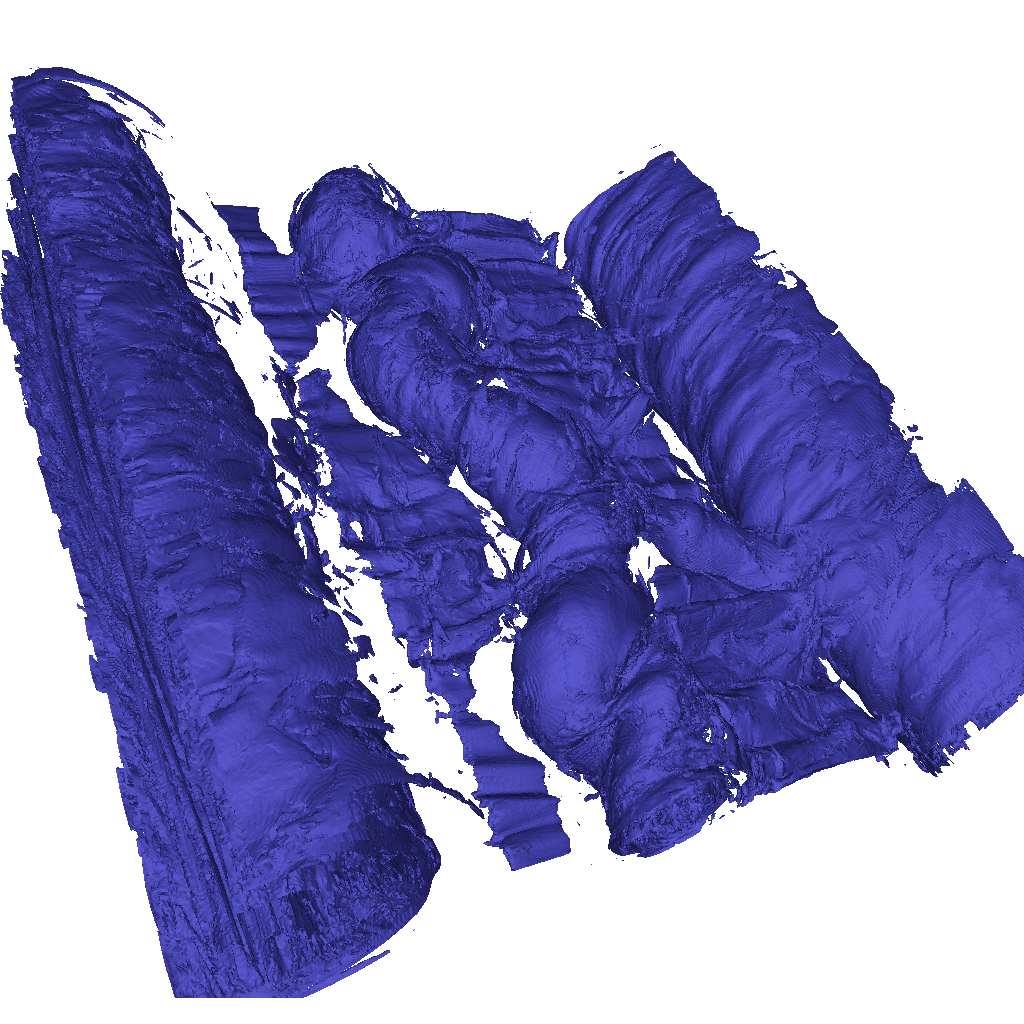}
            & \makecell{512$^\text{3}$ \\ 537MB}
            & \makecell{67MB \\ $\approx\frac{1}{8}$ data size}
            & 1.19 & 15\% & 81MB & 9\% & 9.5M & 229MB \\

        \includegraphics[width=\tableimgwidth]{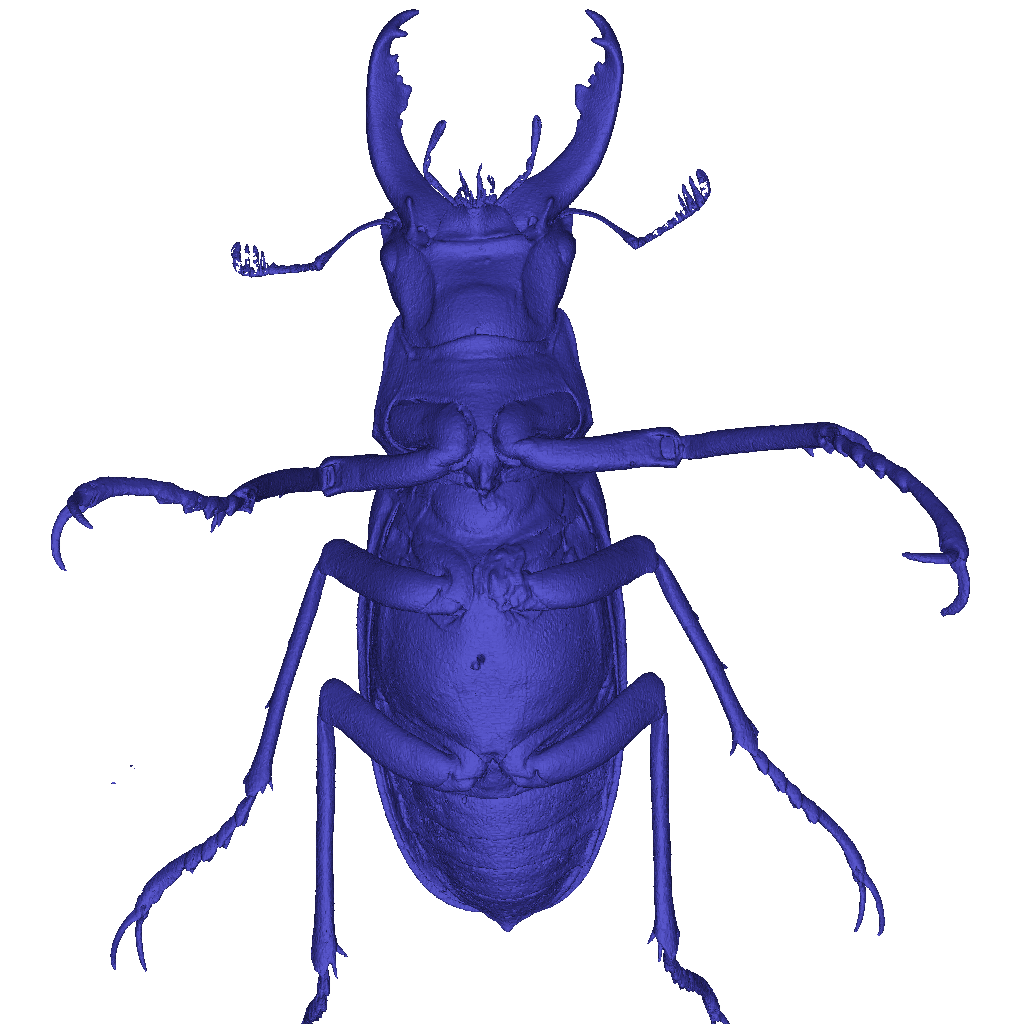}
            & \makecell{832$\times$832$\times$494 \\ 1.37GB}
            & \makecell{86MB \\ $\approx\frac{1}{16}$ data size}
            & 868 & 5\% & 70MB & 3\% & 6.7M & 160MB \\

        \includegraphics[width=\tableimgwidth]{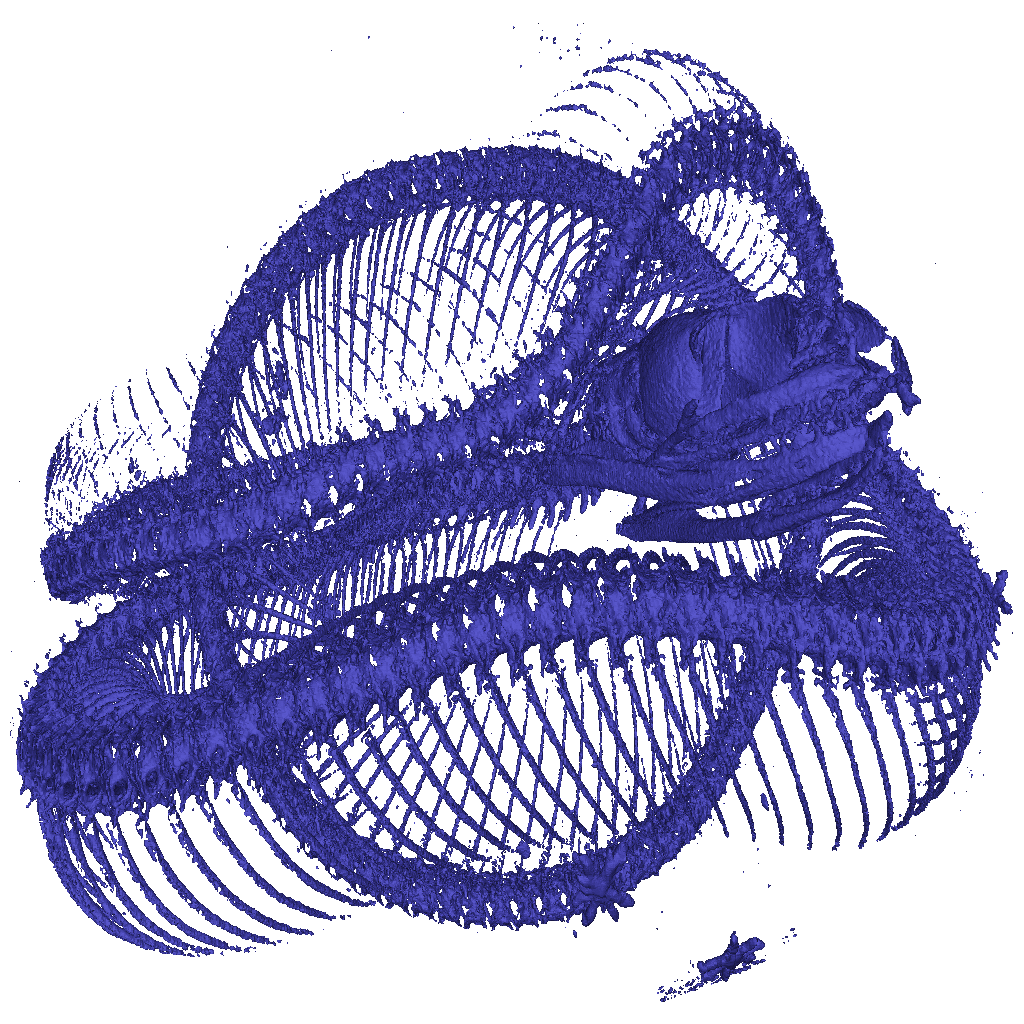}
            & \makecell{1024$\times$1024$\times$795 \\ 3.33GB}
            & \makecell{209MB \\ $\approx\frac{1}{16}$ data size}
            & 111 & 5\% & 167MB & 2\% & 10.2M & 245MB \\

        \includegraphics[width=\tableimgwidth]{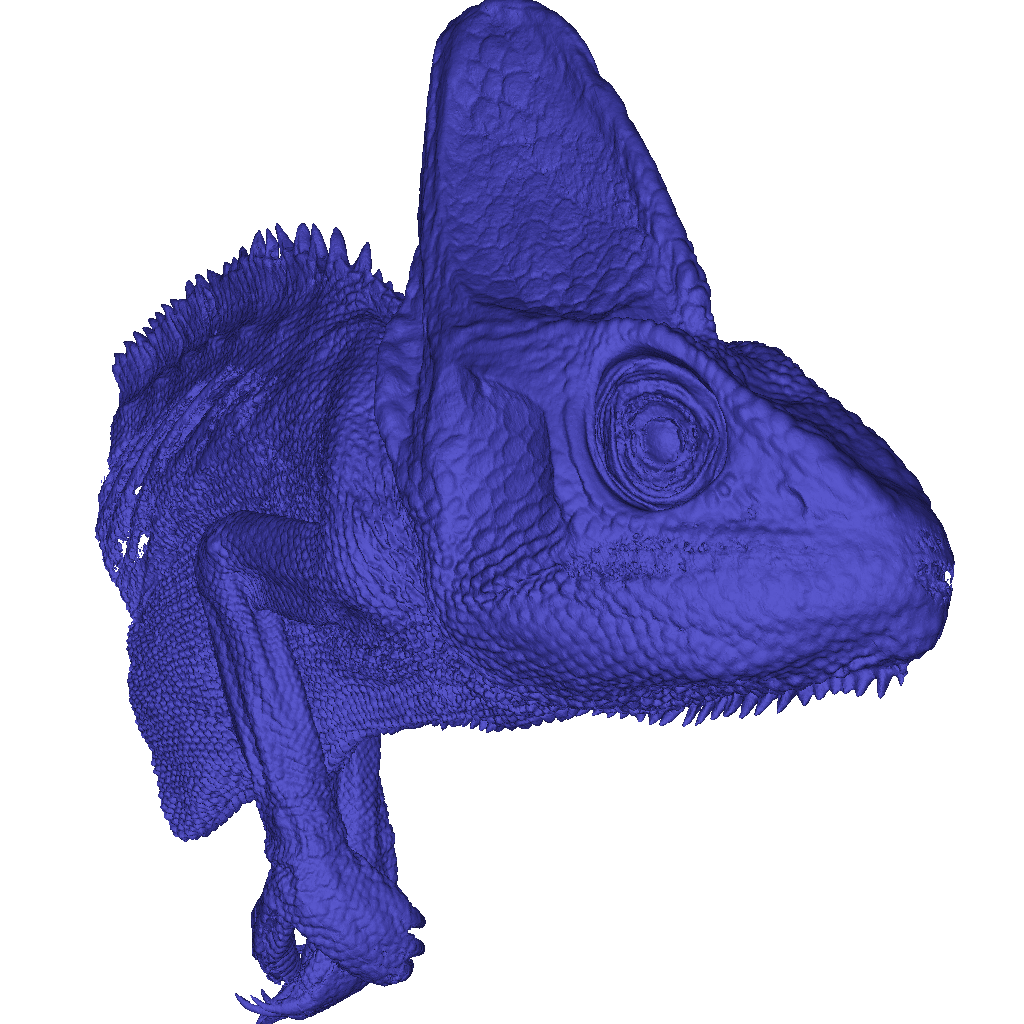}
            & \makecell{1024$\times$1024$\times$1080 \\ 4.53GB}
            & \makecell{283MB \\ $\approx\frac{1}{16}$ data size}
            & 21290 & 4.5\% & 227MB & 1.9\% & 11.2M & 270MB \\

        \includegraphics[width=\tableimgwidth]{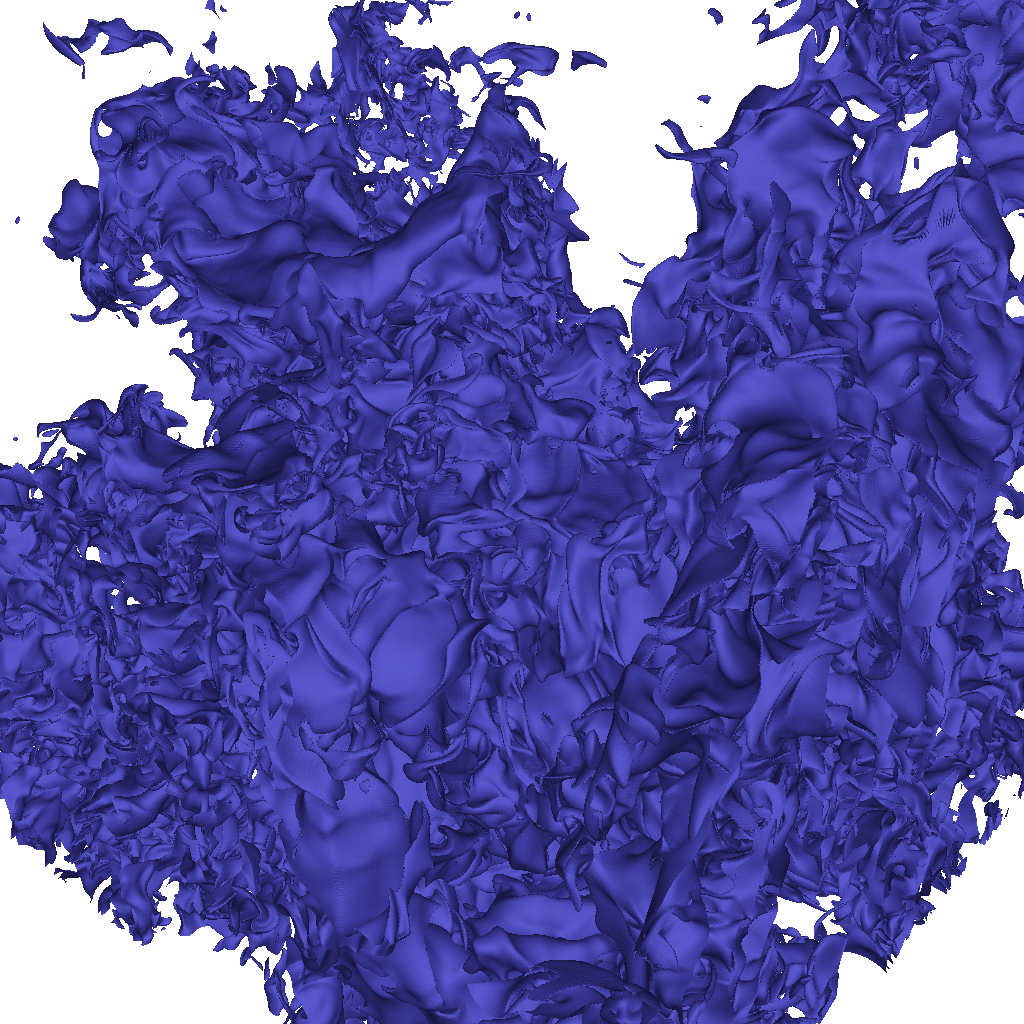}
            & \makecell{1024$^\text{3}$ \\ 4.3GB}
            & \makecell{537MB \\ $\approx\frac{1}{8}$ data size}
            & 1.39 & 26\% & 1.1GB & 12\% & 67M & 1.6GB \\

        \includegraphics[width=\tableimgwidth]{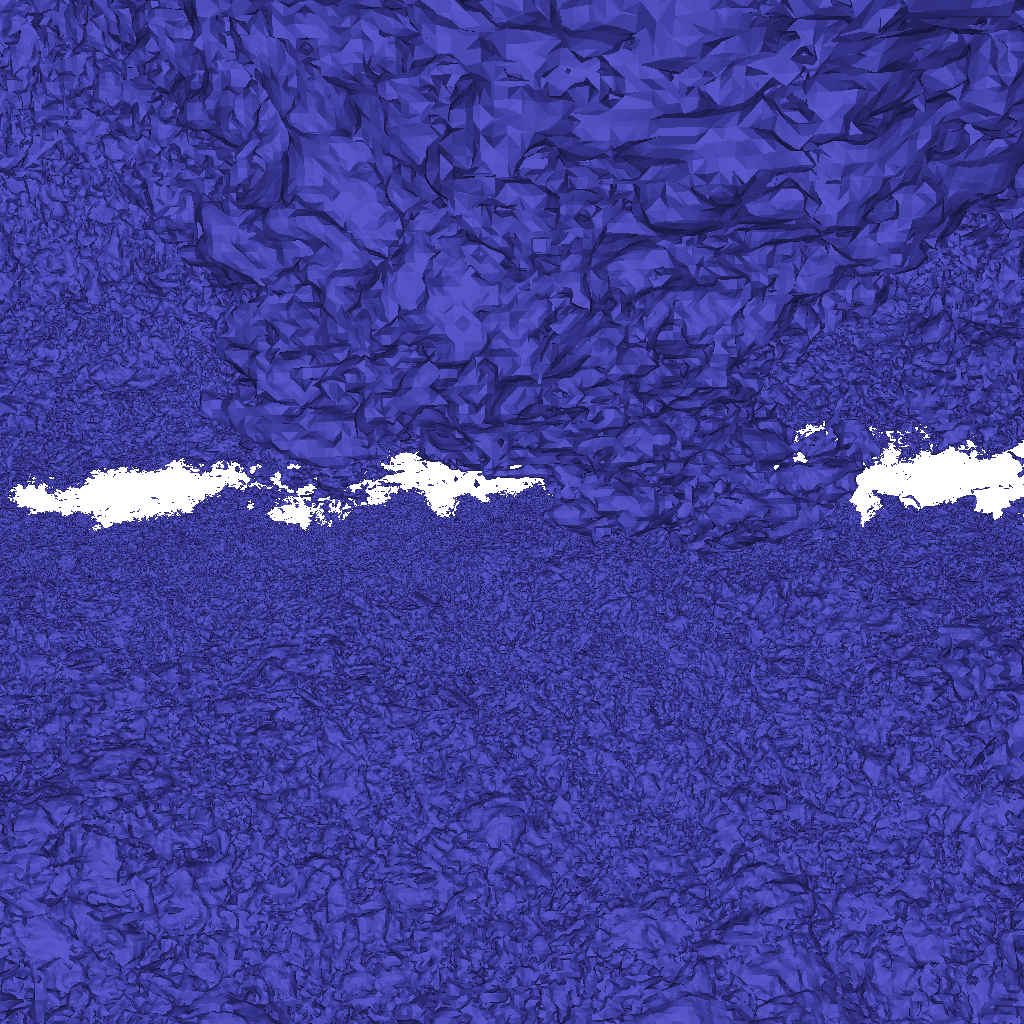}
            & \makecell{10240$\times$7680$\times$1536 \\ 966GB}
            & \makecell{199MB \\ $\approx\frac{1}{4854}$ data size}
            & 0.936 & 43\% & 1.4GB & 24\% & 135M & 3.3GB \\
        \bottomrule
    \end{tabular}
    }
    \vspace{-1em}
    \caption{\label{tab:mc_benchmark_data}%
    Example isosurfaces on the data sets used in our benchmarks. Isosurfaces are typically sparse,
    requiring little data to be decompressed and cached to compute each surface, and even fewer
    blocks to be processed to compute the surface geometry.
    \vspace{-1em}}
\end{table*}

\begin{table*}
    \centering
    \relsize{-1}{
    \begin{tabular}{@{}lrrrrrrrrr@{}}
        \toprule
        & \multicolumn{3}{c}{Random} & \multicolumn{3}{c}{Sweep Up} & \multicolumn{3}{c}{Sweep Down} \\
        \cmidrule(lr){2-4} \cmidrule(lr){5-7} \cmidrule(lr){8-10}
        Dataset & Hit Rate & RTX 2070 & Surface Pro 7 & Hit Rate & RTX 2070 & Surface Pro 7 & Hit Rate & RTX 2070 & Surface Pro 7 \\
        \midrule
        Skull & 98.4\% & 57.7ms & 99.8ms & 100\% & 54.4ms & 85.1ms & 94.8\% & 93.1ms & 161.5ms \\
        Foot & 98.9\% & 57.1ms & 107.9ms & 100\% & 58.4ms & 99.2ms & 97.4\% & 90.1ms & 164.4ms \\
        Backpack & 99.4\% & 68.1ms & 251.0ms & 100\% & 99.1ms & 216.3ms & 96.8\% & 109.7ms & 349.3ms \\
        Plasma & 96.4\% & 107.2ms & 318.6ms & 99.8\% & 94.3ms & 265.3ms & 98.5\% & 126.1ms & 431.8ms \\
        Stag Beetle & 98.2\% & 123.1ms & 322.1ms & 99.6\% & 125.1ms & 337.8ms & 92.0\% & 151.2ms & 464.5ms \\
        Kingsnake & 99.2\% & 212.6ms & 926.5ms & 100\% & 185.3ms & 811.1ms & 97.0\% & 284.5ms & 1306ms \\
        Chameleon & 90.9\% & 371.6ms & 1342ms & 98.9\% & 348.7ms & 1470ms & 96.6\% & 337.1ms & 1482ms \\
        Miranda & 46.8\% & 2218ms & (oom) & 95.7\% & 550.5ms & (oom) & 95.2\% & 549.6ms & (oom) \\
        DNS & 62.9\% & 2632ms & (oom) & 98.4\% & 972ms & (oom) & 97.9\% & 841.3ms & (oom) \\
        \bottomrule
    \end{tabular}
    }
    \vspace{-1em}
    \caption{\label{tab:mc_benchmark_table}%
    Average cache hit rates and isosurface computation performance for the data sets and benchmarks performed.
    Isosurfaces are typically sparse and occupy neighboring regions of the domain, leading to high cache
    rates for the sweep benchmarks. The topology of the nested isosurfaces also allows for high
    hit rates on random isovalues due to the high amount of overlap,
    whereas the turbulent sheet isosurfaces do not overlap as much and see far lower hit rates.
    Although the cache space required for the Miranda and DNS is small enough to fit
    in the Surface Pro's 3.8GB VRAM, for most isovalues, the cumulative size of the cache and output vertex data is not.}
\end{table*}

\subsection{Vertex Computation}
\label{sec:mc_vertex_compute}

Each thread group loads an occupied
block into shared memory and computes the number of vertices that will be output
by each voxel as before, writing the result to shared memory.
The thread group then performs a parallel prefix sum over this shared memory to
compute the per voxel output locations relative to the block's global offset.
Each thread then computes the vertices for its dual cell and writes them to the
global vertex buffer (\Cref{fig:algorithm_sketch}g).
Outputting an indexed triangle mesh
using Flying Edges~\cite{schroeder_flying_2015} within each block,
as proposed by Liu et al.~\cite{liu_parallel_2016}, is also possible.

To reduce the memory required to store the vertex buffer, we adopt a compressed vertex format.
Vertices are stored relative to the block origin and quantized
to 10 bits per coordinate. For each vertex, we store the three 10 bit coordinates packed into a
uint32, along with the block ID as a second uint32, for a total of 8 bytes per vertex.
Blocks are indexed in row-major order, providing a straightforward mapping from
ID to 3D position in the volume.
When rendering, we compute the block position from its ID
and use it to offset the dequantized vertex.

\subsection{Performance Evaluation} 

We evaluate our approach on nine data sets, varying in size
and isosurface topology (\Cref{tab:mc_benchmark_data}).
As ZFP only supports floating point data, we convert the non-float data
sets to 32 bit floating point before compressing them.
ZFP is able to compress the data sets to
far less than 8 or 16 bits per voxel, and bitrates as low as 2
have been demonstrated to have low impact on quality~\cite{lindstrom_fixed_rate_2014}.
Our benchmarks are run on a Surface Pro 7 with an integrated GPU with 3.8GB of VRAM
and a desktop with an RTX~2070 with 8GB of VRAM.

The isosurfaces used in our evaluation can be classified into
``nested'' and ``turbulent sheet'' topologies,
and have different cache behaviors.
Isosurfaces with nested topologies are typical in MRI and CT scans
(Skull, Foot, Backpack, Stag Beetle, Kingsnake, and Chameleon),
although they do occur in simulations as well (Plasma).
In the nested topology isosurfaces, surfaces at lower values enclose those
at higher values.
Such isosurfaces can be extremely cache friendly,
as after computing a surface at a lower value, a large number of the blocks
will be reused for neighboring surfaces, and even distant enclosed ones.

Isosurfaces with turbulent sheet topologies are typically found in simulations
where the surface represents a moving interface, e.g., fluids mixing or turbulent
flows (Miranda and DNS).
In a turbulent sheet topology isosurface, surfaces move along one or more axes of the
domain with the isovalue, and do not enclose each other. Moreover, the turbulent
nature of the surface results in a large number of blocks being occupied and
output triangles.
Isosurfaces with turbulent sheet topology tend to be less cache friendly for random isovalues, as different
surfaces occupy different and distant regions of the domain,
with relatively few shared blocks; however, neighboring isovalues do share blocks
as the surface does not make large jumps in the domain.



\subsubsection{Isosurface Extraction}
%
%

We perform three benchmarks on each data set, covering different
levels of cache friendliness and user interaction modes when exploring a volume.
The first benchmark computes 100 random isovalues, as in~\Cref{sec:example_mc_perf},
representing a cache-unfriendly exploratory use case.
The other two benchmarks
sweep the isosurface up or down the data set's value range,
as may be done when comparing neighboring values in the data,
and are relatively cache friendly.

Although the sweep up and down
benchmarks may seem to be the same, the cache behavior of the
two differ on the nested isosurfaces.
On these data sets, an up sweep will frequently operate in cache,
as the number of active blocks decreases and the inner surfaces are likely to
require blocks that were already decompressed for the previous containing surface.
However, a down sweep
is less likely to hit cache, as the surface area of the isosurface increases
to cover more blocks, with these blocks less likely to be cached.

As before, the isovalues are sampled over a value range covering surfaces
typically of interest to users, excluding noise and values so high as
to produce few output triangles. Example configurations for each data set
are shown in~\Cref{tab:mc_benchmark_data}.
In each benchmark, we discard the first computation as we have found WebGPU to
have a high first launch overhead.
The average cache hit rates and extraction time
for the benchmarks are shown in~\Cref{tab:mc_benchmark_table}.

On the nested isosurface data sets,
we observe high cache hit rates for all three benchmarks,
with Sweep Down performing
the worst on average due to the lower cache hit rate. A lower cache hit rate
means a larger number of blocks must be decompressed to compute each
surface, impacting compute time. It is interesting to note that
Random achieves a higher hit rate than Sweep Down.
After computing some sampling of the nested isosurfaces in Random,
the cache contains the most frequently used blocks across
these surfaces, allowing for high cache hit rates and
better performance.

In contrast, on the turbulent sheet data sets, we observe high cache
hit rates on the Sweep benchmarks and low hit rates on Random.
When computing random isovalues, the surface can make large jumps
in the domain, covering entirely different regions of the volume.
This issue is exacerbated by the high resolution of the data sets and the turbulent
nature of the computed isosurfaces, leading to a large number of blocks
covered by each surface with few shared between them. As a result, large numbers
of blocks must be decompressed each time, severely impacting compute time.
However, the surface does not move significantly when sweeping the isovalue up
or down, resulting in high cache hit rates and better performance.

As indicated by the cache hit rates, we find that on the nested
isosurface data sets so few new blocks must be decompressed to
compute each new surface (on average, $< 0.5\%$ of the total blocks
in the data set)
that decompression does not occupy a large portion of compute time.
The bulk of time is spent in the first pass to select the active
blocks and the mark new items step of the cache update,
the latter of which then determines no new items are to be added.

In contrast, on the turbulent sheet data sets a larger percentage
of time is spent in decompression due to the higher miss rate and
higher absolute number of active blocks. On the random benchmarks,
decompression occupies an average of 63\% of compute time, with
an average of 12\% of blocks decompressed for each new surface.
On the sweep benchmarks, time is more evenly divided among
decompression, finding active blocks, and updating the cache,
with an average of 0.8\% of blocks decompressed for each new surface.

We find that when new items need to be added to the cache,
the bulk of the cache update time is spent sorting the available slots by age.
In WebGPU we do not have access to optimized libraries such as Thrust~\cite{bell_thrust_2017}
and VTK-m~\cite{moreland_vtkm_2016}, and our sort by key implementation becomes a bottleneck.

\subsubsection{Rendering Performance}
We report the framerate for the example isosurfaces computed in~\Cref{tab:mc_benchmark_data}
in~\Cref{tab:mc_rendering_perf}. The output of our isosurface computation
is a triangle soup in our compressed vertex format, which we render with WebGPU.
On both the Surface Pro 7 and RTX 2070, we
achieve real-time rendering performance, even for large isosurfaces.
Although our rendering modality is relatively simple, these results are
encouraging for large-scale data visualization in general, demonstrating
that WebGPU can achieve high framerates for large geometry.
The quality of the isosurfaces are primarily dependent on the reconstruction
accuracy of the compression method, which have demonstrated
high accuracy even at high compression ratios~\cite{lindstrom_fixed_rate_2014,hoang-precision-tvcg-2019,hoang_efficient_2021}.

We note that, as the vertex buffer occupies a substantial amount of memory,
it would be valuable to explore applying implicit isosurface raycasting
methods on top of the existing block structure. The blocks can be
seen as a macrocell grid~\cite{parker_interactive_1998} for space-skipping,
and implicit isosurface ray tracing performed within these blocks.

\begin{table}
    \centering
    \relsize{-1}{
    \begin{tabular}{@{}lrrr@{}}
        \toprule
        Dataset & Triangles & RTX 2070 (FPS) & Surface Pro 7 (FPS) \\
        \midrule
        Skull & 2.1M & 180 & 44 \\
        Foot & 743K & 174 & 89 \\
        Backpack & 6.2M & 146 & 25 \\
        Plasma & 9.5M & 118 & 24 \\
        Stag Beetle & 6.7M & 139 & 19 \\
        Kingsnake & 10.2M & 128 & 24 \\
        Chameleon & 11.2M & 123 & 19 \\
        Miranda & 67M & 53 & (oom) \\
        DNS & 135M & 36 & (oom) \\
        \bottomrule
    \end{tabular}
    }
    \vspace{-1em}
    \caption{\label{tab:mc_rendering_perf}%
    Rendering performance for the isosurfaces shown in~\Cref{tab:mc_benchmark_data}.
    We find that WebGPU is capable of interactive rendering of large triangle meshes
    even on lightweight clients.
    \vspace{-2em}}
\end{table}

\section{Conclusion: A Roadmap for Client-side Scientific Visualization in the Browser}

We have conducted an experimental assessment of the suitability
of the browser for building scalable scientific visualization
applications, and conclude that by leveraging WebAssembly
and WebGPU, it is possible to deliver powerful visualization
applications that run entirely in the client's browser.
We have shown that native libraries can be
compiled to WebAssembly to deploy them in the browser without
a significant porting effort or loss of performance.
Moreover, we have found that WebGPU's modern low overhead
rendering and compute API can achieve performance on par
with native Vulkan implementations.
With our GPU parallel isosurface extraction algorithm for
block-compressed data, we have demonstrated that interactive
isosurface computation on massive data sets can be
achieved entirely in the browser by making the appropriate
algorithm design choices and precision and resolution trade-offs.
We hope that this work can serve as a call to action and motivate
further efforts to bring scientific visualization to client-side
browser applications.


The biggest challenge faced when bringing scientific visualization
to the web is the lack of available libraries compiled
to WebAssembly or leveraging WebGPU. Having researchers or developers track down the
potentially large chain of dependencies required to compile
their needed libraries to WebAssembly or implement the complex GPU parallel
primitives needed by their algorithm is clearly undesirable.
Providing packages with generated bindings for
widely used libraries, such as VTK, would be valuable.
A package providing optimized WebGPU implementations
of data-parallel primitives similar to Thrust~\cite{bell_thrust_2017}
and VTK-m~\cite{moreland_vtkm_2016},
would be critical to ease development of applications
that process large data sets.

\acknowledgments{
The authors wish to thank Duong Hoang for providing compressed versions of the DNS
and Stefan Zellmann for valuable discussions.
This work was funded in part by NSF OAC awards 1842042, 1941085,
NSF CMMI awards 1629660, LLNL LDRD project SI-20-001.
This material is based in part upon work supported by the Department of Energy,
National Nuclear Security Administration (NNSA), under award DE-NA0002375.
This research was supported in part by the Exascale Computing Project (17-SC-20-SC),
a collaborative effort of the U.S. Department of Energy Office of Science and the NNSA.
This work was performed in part under the auspices of the U.S. Department of Energy by Lawrence Livermore National
Laboratory under Contract DE-AC52-07NA27344.}

\bibliographystyle{abbrv-doi}

\bibliography{webscivis}
\end{document}